\documentclass[journal]{IEEEtran}
\usepackage{amssymb}
\usepackage{amsfonts}
\usepackage{amsmath}
\usepackage{algorithm}
\usepackage{algorithmic}
\usepackage{cite}
\usepackage{stfloats}
\usepackage{amsmath}
\usepackage{graphicx}
\usepackage{amsfonts,amsmath,amssymb}
\usepackage{url}
\usepackage{bm}
\usepackage{bbm}
\usepackage{subfigure}
\usepackage{algorithmic}
\usepackage{algorithm}
\usepackage{color}
\newcommand{\Hnull}{\mathcal{H}_0}
\newcommand{\Halt}{\mathcal{H}_1}

\newcommand{\Honull}{\mathcal{{D}}_0}
\newcommand{\Hoalt}{\mathcal{{D}}_1}

\newtheorem{proposition}{\textbf{Proposition}}
\newtheorem{lemma}{\textbf{Lemma}}

\usepackage{amssymb}
\usepackage{}
\usepackage{amsfonts}
\usepackage{amsmath}
\usepackage{algorithm}
\usepackage{algorithmic}
\usepackage{cite}

\usepackage{graphicx,subfigure,amsmath,amssymb,cite}
\usepackage{float}
\ifCLASSINFOpdf
\else
\fi
\begin{document}
\title{Joint Optimization of a UAV's Trajectory and Transmit Power for Covert Communications}

\author{\IEEEauthorblockN{Xiaobo Zhou, Shihao Yan, \IEEEmembership{Member, IEEE,}  Jinsong Hu, \IEEEmembership{Member, IEEE,}\\ Jiande Sun, \IEEEmembership{Member, IEEE,} Jun Li, \IEEEmembership{Senior Member, IEEE,} and Feng Shu, \IEEEmembership{Member, IEEE}}\\

\thanks{X. Zhou is with the School of Physics and Electronic Engineering, Fuyang Normal University, Fuyang 236037, China. X. Zhou is also with the School of Nanjing University of Science and Technology, Nanjing 210094, China (e-mail: zxb@fync.edu.cn).}
\thanks{S. Yan is with the School of Engineering, Macquarie University, Sydney, NSW 2109, Australia (e-mail: shihao.yan@mq.edu.au).}
\thanks{J. Hu is with the College of Physics and Information, Fuzhou University, Fuzhou 350116, China (e-mail: jinsong.hu@fzu.edu.cn).}
\thanks{J. Sun is with the School of Information Science and Engineering, Shandong Normal University, Jinan 250358, China (e-mail: jiandesun@sdnu.edu.cn).}
\thanks{J. Li and F. Shu are with the School of Electronic and Optical Engineering, Nanjing University of Science and Technology, Nanjing 210094, China (e-mails: \{jun.li, shufeng\}@njust.edu.cn).}
}

\maketitle

\begin{abstract}
This work considers covert communications in the context of unmanned aerial vehicle (UAV) networks, aiming to hide a UAV for transmitting critical information out of an area that is monitored and where communication is not allowed. Specifically, the UAV as a transmitter intends  to transmit information to a legitimate receiver (Bob) covertly while avoiding being detected by a warden (Willie), where location uncertainty exists at Bob and/or Willie.
In order to enhance the considered covert communication performance, we jointly optimize the UAV's trajectory and transmit power in terms of maximizing the average covert transmission rate from the UAV to Bob subject to transmission outage constraint and covertness constraint. The formulated optimization problem is difficult to tackle directly due to the intractable constraints. As such, we first employ conservative approximation to transform a constraint into a deterministic form and then apply the first-order restrictive approximation
to transform the optimization problem into a convex form. By applying the successive convex approximation (SCA) technique, an efficient iterative algorithm is developed to solve the optimization problem. Our examination shows that the developed joint trajectory and transmit power optimization scheme achieves significantly better covert communication performance as compared to a benchmark scheme.
\end{abstract}

\IEEEpeerreviewmaketitle

\section{Introduction}

Recently, unmanned aerial vehicles (UAVs) have been extensively used in wireless communication networks, due to their controllable mobility, on-demand deployment, and line-of-sight (LoS) air-to-ground link (e.g., \cite{Zeng2016Wireless}). For example, a UAV can be used as an airborne wireless communication platform such as mobile base station (BS) to rapidly recover communication service or to enhance communication quality (e.g., \cite{TrajZeng2018,Wu2018Common,Wu2018Capacity,JointWu2018,UnmanHWANG2018}). It can also be utilized as a mobile relay to provide a wireless connection between two or more remote users without the need for a reliable direct communication link (e.g., \cite{Zeng2016throughput,Zhang2018Joint,UAV2018Xiao}). Furthermore, a UAV can conduct mobile data collection or information dissemination to assist with various Internet of Things (IoT) applications (e.g., \cite{Zhan2018Energy,Lyu2016Cyclical}).

In the aforementioned UAV applications, wireless communication security is of increasing concerns due to the broadcast nature of wireless channels. The LoS air-to-ground link in the UAV networks makes the confidential messages transmitted by a UAV being easier to be intercepted by illegal eavesdroppers on the ground. Against this background, several recent works addressed wireless communication security in the context of the UAV networks (e.g.,\cite{Zhang2017Securing,Li2018Enabled,Zhou2018Improving,Cai2018Dual,Wang2017Improving,Zhao2018Caching}).
In general,
the security performance of UAV networks can be improved by adjusting the UAV's flight trajectory and transmit power to simultaneously increase
the channel quality of legitimate links and degrade the channel quality of eavesdropping links.
For example, in \cite{Zhang2017Securing}, a UAV communicates with a ground user and prevents the confidential messages from being intercepted by the eavesdropper via optimizing its flight trajectory and transmit power. Meanwhile, a UAV-enabled mobile jammer was considered in \cite{Li2018Enabled,Zhou2018Improving} to enhance wireless communication security in the context of UAV networks, where the friendly UAV jammer was employed to transmit artificial noise (AN) to improve the communication security. In \cite{Cai2018Dual}, the authors considered a scenario of dual UAVs, where one UAV transmitted the confidential messages to ground users and the other UAV transmitted AN to interfere with the eavesdroppers. The secure communication in a mobile relay network was considered in \cite{Wang2017Improving}, where a UAV acted as a mobile relay to enhance physical layer security by dynamically adjusting its location and transmit power. Furthermore, a caching UAV assisted secure transmission in hyper-dense networks based on interference alignment was investigated in~\cite{Zhao2018Caching}, where UAVs were employed to provide data traffic to users, aiming to reduce the transmission pressure of small-cell base stations.

The aforementioned works in the context of UAV network security only focused on preventing the confidential information from being intercepted by an eavesdropper. However, hiding the wireless transmission of a UAV or the UAV itself has been overlooked, which is a critical issue in the context of UAV networks.
For example, in military applications once the enemy detects the transmission behavior of a UAV (such that to detect the UAV itself), it may cause the UAV's location information to be exposed, which makes the UAV vulnerable to attack. Fortunately, we find that the emerging covert communication technology can meet the requirement of hiding the UAV's transmission or the UAV itself, since in covert communications a transmitter intends to transmit information to a receiver covertly in order to avoid this transmission being detected by a warden~\cite{Bash2013Limits,goeckel2016covert,Sobers2017Covert,Hu2018Covert,HeB2018Covert,yan2018gaussian,Shihao2018Delay}. In the context of covert communications, the authors of \cite{Bash2013Limits} proved that transmitting more than $\mathcal{O}(\sqrt{n})$ bits would result in the warden's detection error probability approaching zero. Meanwhile, \cite{goeckel2016covert} showed that the transmitter can covertly transmit $\mathcal{O}(n)$ bits to the receiver when the warden does not exactly know its noise power. In addition to noise power uncertainty, the impact of uninformed jammers on covert communications was examined in \cite{Sobers2017Covert}, where a positive covert rate was proved to be achievable.
Furthermore, covert communications with a poisson field of interferers and within one-way relay networks were investigated in \cite{Hu2018Covert} and \cite{HeB2018Covert}, respectively. Most recently, the impact a finite number of channel uses on covert communications over additive white Gaussian noise (AWGN) channels was considered in \cite{Shihao2018Delay}, where the authors first proved that the optimal actual number of channel uses is the maximum available number of channel uses and then the authors verified that the uniformly distributed random transmit power can be used to further enhance the covert communication performance in delay-constrained application scenarios.
We note that the aforementioned works on covert communications all focused on static scenarios, i.e., the transmitter, the receiver, and the warden are fixed at specific locations, which severely limits the applications of covert communications, especially in military scenarios.

\begin{figure}[!t]
  \centering
  \includegraphics[scale=1.7]{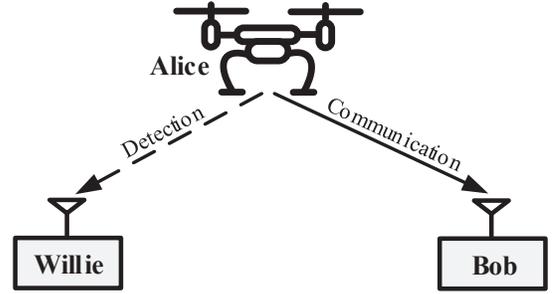}\\
  \caption{Covert communications in the context of UAV networks.}\label{Sys_Sch}
\end{figure}

In order to address the UAV communication covertness (with the ultimate goal of hiding a UAV) and to extend the application scenarios of covert communications,  in this work, we consider covert communications in the context of UAV networks as shown in Fig.~\ref{Sys_Sch}, where the transmitter Alice (UAV) as a mobile transmitter intends to communicate with the ground receiver Bob covertly, while the ground warden Willie wants to detect the UAV's wireless transmission. We note that the UAV as a mobile transmitter can significantly extend the transmission range of covert communications, which is especially useful for military air-to-ground covert communications, while the traditional static scenario for covert communications serves as a special case of this dynamic scenario. Our goal is to maximize the covert transmission rate from the UAV to the legitimate ground user via optimizing the UAV's trajectory and transmit power, which is a new design framework that jointly considers the average covert transmission rate (ACTR) and the average minimum total error rate at Willie. The main contributions of this work are summarized as below.
\begin{itemize}
\item For the first time, we consider a UAV as the transmitter Alice (a mobile transmitter) in the context of covert communications, aiming to extend the applications of covert communications from static scenarios to dynamic ones. In addition, we study a practical setup, where the locations of both the legitimate receiver Bob and the warden Willie are subject to uncertainties. This leads to the fact that the distances from the UAV to both Bob and Willie are random variables following noncentral chi-square distributions, which (as we proved) can be well approximated by Gaussian distributions. Based on this approximation, we derive a lower bound on the average minimum total error rate over Willie's location uncertainty from the perspective of UAV, which enables us to analyze the covert communication performance in the considered scenario.

\item
In order to jointly design the UAV's trajectory and transmit power, we formulate an optimization problem that maximizes the ACTR subject to the transmission outage probability constraint at Bob and covertness constraint at Willie as well as the UAV's mobility constraint and transmit power constraint. Such a joint optimization problem is generally difficult to tackle directly due to the non-convex constraints. To solve this optimization problem, we first transform the intractable transmission outage probability constraint into a deterministic form by using the conservative approximation and then we apply the first-order restrictive approximation to transform the optimization problem into a convex form, which is mathematically tractable.

\item
We employ the successive convex approximation (SCA) algorithm to solve the achieved convex optimization problem iteratively and present a method that generates initial feasible solutions, which outputs the UAV's trajectory and transmit power for covert communications in the considered scenario.
Our examination shows that the proposed joint trajectory and transmit power optimization scheme achieves significant covert communication performance gain relative to a benchmark scheme. Interestingly, our examination also shows that, with regard to the UAV's trajectory, it prefers to hover around Bob for a certain period and, surprisingly, it moves closer to Willie as the covertness becomes stricter.
\end{itemize}

The reminder of this work is organized as follows. In Section II, we present the considered system model. In Section III, we derive the minimum total error rate at Willie and average minimum total error rate at Willie from the perspective of the UAV. In Section IV, we joint design the trajectory and transmit power of the UAV to maximize the ACTR. Section V provides our numerical results to examine the covert communication performance of the proposed scheme relative to the benchmark scheme and Section VI draws conclusion remarks.

$\emph{Notation}$: Boldface lowercase and uppercase letters represent vectors and matrices, respectively, $\mathbf{A}^T$, $\mathbf{A}^H$, and $\mathbf{A}\succeq\mathbf{0}$ denote  transpose, conjugate transpose, and semidefiniteness of matrix $\mathrm{\mathbf{A}}$, respectively. $\mathbb{E}_x[\cdot]$, $\mathrm{Tr}(\cdot)$, $\|\cdot\|$, and $|\cdot|$ denote the statistical expectation of $x$, trace, Euclidean norm, and absolute value, respectively.
$\mathrm{Pr}\{\cdot\}$ and $\dot{\mathbf{x}}(t)$ denote the probability of an event and the derivative with respect to $t$, respectively.
$\mathbb{N}(0,\sigma^2)$ denotes the Gaussian distribution with zero mean and variance $\sigma^2$.

\section{System Model}

\subsection{Considered Scenario and Adopted Assumptions}
As shown in Fig.~\ref{Sys_Sch}, in this work we consider covert communications in the context of UAV networks, in which Bob and Willie are located on the ground, while Alice as a UAV aims to covertly communicate with Bob without being detected by Willie. We assume that each of UAV, Bob, and Willie is equipped with a single antenna. Without loss of generality, we use a three-dimensional Cartesian coordinate system to indicate the location information of each transceiver. As such, we denote the horizontal location of Bob and Willie by $\mathbf{q}_b\triangleq[x_b,y_b]^T\in\mathbb{R}^{2\times 1}$ and $\mathbf{q}_w\triangleq[x_w,y_w]^T\in\mathbb{R}^{2\times 1}$, respectively.
We assume that UAV flies at a fixed altitude above the ground, denoted by $H$, which can be considered as the minimum altitude to avoid collision with ground obstacles. The flight period and the maximum flying speed of the UAV are set as $T$ and $V_{\max}$, respectively. We note that the takeoff and the landing locations of the UAV are usually \textit{a priori} given according to its specific task. We denote $\mathbf{q}_{a,0}\triangleq[x_{a,0},y_{a,0}]^T$ and $\mathbf{q}_{a,F}\triangleq[x_{a,F},y_{a,F}]^T$ as the initial and final locations of UAV, respectively.
At time instant $t$, the horizontal coordinate of the UAV is denoted as $\mathbf{q}_a(t)\triangleq[x_a(t),y_a(t)]^T\in\mathbb{R}^{2\times 1}$, where $0\leq t\leq T$.
As such, we have the initial and final location constraints as given by $\mathbf{q}_a(0)=\mathbf{q}_{a,0}$ and $\mathbf{q}_a(T)=\mathbf{q}_{a,F}$, while the flying speed constraint is given by $\mathbf{\dot{q}}_a(t)\leq V_{\max}$, $0\leq t\leq T$, where $\dot{\mathbf{q}}_a(t)$ denotes the instantaneous velocity of UAV,  which is the time-derivative of $\mathbf{q}(t)$.

In UAV networks, continuous time $t$ implies an infinite number of speed constraints, which makes the trajectory design being too hard to tackle. Therefore, we equally divide the flight period $T$ into $N$ time slots. The period of each time slot $\delta_t=\frac{T}{N}$ is set to be sufficiently small such that the location of the UAV is considered to be approximately unchanged within each time slot.
As such, UAV's flight trajectory can be expressed as $\mathbf{q}_a[n]\triangleq[x_a[n],y_a[n]]^T\in\mathbb{R}^{2\times 1}$, and $n=\{1,\cdots,N\}\triangleq\mathcal{N}$. We note that infinitesimal $\delta_t$ can accurately approximate the continuous flight trajectory of UAV. However, it also leads to high computational complexity. Therefore, the time slot $\delta_t$ should be properly chosen to balance between the approximation accuracy and the computational complexity in UAV networks~{{\cite{JointWu2018}}}.
The maximum flying distance of UAV at each time slot, denoted by $L$, can be expressed as $V_{\max}\delta_t$. According to the above description, we write the mobility constraint on UAV as
\begin{subequations}\label{Mob_C}
\begin{align}
&\|\mathbf{q}_a[n+1]-\mathbf{q}_a[n]\|\leq L,n\in\mathcal{N}\setminus\{N\},\label{Mob_Ca}\\
&\|\mathbf{q}_a[1]-\mathbf{q}_{a,0}\|\leq L,~\mathbf{q}_a[N]=\mathbf{q}_{a,F}.\label{Mob_Cb}
\end{align}
\end{subequations}
We use $P_a[n]$ to denote the transmit power of the UAV at the $n$-th time slot, while $\bar{P}_{a,\max}$ and $P_{a,\max}$ denote the maximum average transmit power and peak transmit power of UAV, respectively. Thus, the average and the peak transmit power constraints at UAV are given by
\begin{subequations}\label{Pow_C}
\begin{align}
&\frac{1}{N}\sum_{n=1}^NP_a[n]\leq \bar{P}_{a,\max},\label{Pow_Ca}\\
&0\leq P_a[n]\leq P_{a,\max},~\forall n,\label{Pow_Cb}
\end{align}
\end{subequations}
respectively.
We note that the maximum average transmit power $\bar{P}_{a,\max}$ is to keep the long-term power budget of the communication related power.

\subsection{Channel Model}

It is assumed that the channels from  UAV to Bob and to Willie are all dominated by LoS components~\cite{JointWu2018}. As such, UAV has to estimate the locations of Willie and Bob in order to determine the corresponding channels.
In practice, location estimation suffers from estimation errors. In this work, we assume that the UAV knows the potential regions in which Willie
is located while the exact location of Willie is unknown, i.e., Willie's location is assumed to be known by UAV with some uncertainty (e.g., localization errors). This assumption can be justified through the following two aspects.
Firstly, the UAV can use a camera or radar to detect suspicious areas where Willie may exist, which was clarified in \cite{Li2018Enabled}.
Secondly, this assumption can be justified by considering the scenarios where Willie was a licensed user who had legal access to the UAV networks (and thus the UAV knows his imperfect location information based on his previous communications with the UAV networks), but the UAV suspects Willie as a malicious user for the current transmission to Bob (i.e., the UAV intends to hide the transmission to Bob from Willie). We note that similar scenarios have been widely considered in the literature of physical layer security (e.g., \cite{Venkaya2017On,Hama2018Class}).
In addition, a widely used Gaussian error model is used model the location errors of Bob and Willie, which leads to that the location coordinates of Bob and Willie are given as
\begin{subequations}\label{E_Loc}
\begin{align}
x_b&=\hat{x}_b+\Delta x_b,~y_b=\hat{y}_b+\Delta y_b,\label{E_Loca}\\
x_w&=\hat{x}_w+\Delta x_w,~y_w=\hat{y}_w+\Delta y_w,\label{E_Locb}
\end{align}
\end{subequations}
where $\mathbf{\hat{q}}_b\triangleq[\hat{x}_b, \hat{y}_b]^T\in\mathbb{R}^{2\times 1}$ and $\mathbf{\hat{q}}_w\triangleq[\hat{x}_w, \hat{y}_w]^T\in\mathbb{R}^{2\times 1}$ are estimated horizontal coordinates of Bob and Willie, respectively, while $\mathbf{e}_b\triangleq[\Delta x_b,\Delta y_b]^T\in\mathbb{R}^{2\times 1}$ and $\mathbf{e}_w\triangleq[\Delta x_w,\Delta y_w]^T\in\mathbb{R}^{2\times 1}$ are the estimation errors on the locations of Bob and Willie, respectively. We note that $\Delta x_b$ and $\Delta y_b$ are independent and identically distributed (i.i.d.) Gaussian random variables, each of which follows  $\mathbb{N}(0,\varepsilon_b^2)$, i.e., $\mathbf{e}_b\sim\mathbb{N}(\mathbf{0},\varepsilon_b^2\mathbf{I}_2)$. Meanwhile, each of $\Delta x_w$ and $\Delta y_w$ follows $\mathbb{N}(0,\varepsilon_w^2)$, i.e., $\mathbf{e}_w\sim\mathbb{N}(\mathbf{0},\varepsilon_w^2\mathbf{I}_2)$.
Following \eqref{E_Loc}, at the $n$-th time slot, the channel power gain from UAV to Bob and from UAV to Willie are given by
\begin{subequations}\label{Sys2}
\begin{align}
|h_{ab}[n]|^2&=\frac{\beta_0}{\|\mathbf{q}_a[n]-\mathbf{\hat{q}}_b-\mathbf{e}_b\|^2+H^2},~\forall n,\label{Sys2a}\\
|h_{aw}[n]|^2&=\frac{\beta_0}{\|\mathbf{q}_a[n]-\mathbf{\hat{q}}_w-\mathbf{e}_w\|^2+H^2},~\forall n,\label{Sys2b}
\end{align}
\end{subequations}
respectively, where $\beta_0$ denotes the channel power gain at the reference distance $1~\mathrm{m}$.


\subsection{Binary Hypothesis Testing at Willie}
In covert communications, Willie has to decide whether UAV transmitted to Bob. We note that, since UAV flies over the area above Willie, i.e., Willie can directly observe UAV's flight trajectory, it is reasonable to assume that the Willie has the perfect knowledge of UAV's location information. This also can be justified by the fact that this is the best-case scenario for Wille (i.e., the worst-case scenario for covert communications).
As such, for the $i$-th channel use in the $n$-th time slot, the received signal at Willie is given by
\begin{align}\label{Hy_test}
y_w^i[n]=
\begin{cases}
n_w(i),&\Hnull,\\
\sqrt{\frac{\beta_0 P_a[n]}{\|\mathbf{q}_a[n]-\mathbf{q}_w\|^2+H^2}}x(i)+n_w(i),&\Halt,
\end{cases}
\end{align}
where $\Hnull$ denotes the null hypothesis in which  UAV did not transmit, and $\Halt$ denotes the alternative hypothesis where UAV did transmit to Bob. $x(i)$,  $i = 1, 2, \dots, m$, represents the signal transmitted by UAV, satisfying $\mathbb{E}[x(i)x(i)^*]=1$, where $m$ is the total number of channel uses in the $n$-th time slot. $y_w^i[n]$ denotes the received signal at Willie for the $i$-th channel use, and $n_w(i)$ is the AWGN at Willie with variance $\sigma_w^2$.
We note that the transmit power $P_a[n]$ is considered as fixed within each time slot, but may change from one time slot to another time slot.

At the $n$-th time slot, the optimal decision rule for Willie that minimizes the total error rate is given by~\cite{Shihao2018Delay}
\begin{align}\label{Like_Test}
T_w[n]\triangleq \frac{1}{m}\sum_{i=1}^m|y_w^i[n]|^2\mathop{\gtreqless}\limits_{\Honull}^{\Hoalt} P_{th}[n],
\end{align}
where $T_w[n]$ and $P_{th}[n]$ are the average power of each received symbol at Willie and the detection  threshold at the $n$-th time slot, respectively, while $\Honull$ and $\Hoalt$ are the decisions in favor of $\Hnull$ and $\Halt$, respectively.
In this work, we consider an infinite number of channel uses in each time slot, i.e., $m\rightarrow\infty$, which can be justified as follows. Firstly, to balance the approximate accuracy and computational complexity, $\delta_t$ can be chosen such that $\frac{V_{\max}\delta_t}{H}\leq \kappa$, where $\kappa$ is a given threshold to control the approximation accuracy \cite{JointWu2018,wu2019safeguarding}. Following this consideration,
we can increase the flying altitude $H$ or reduce the maximum flying speed $V_{\max}$ to appropriately increase the duration of the time slot $\delta_t$. Secondly, we recall that the number of symbols transmitted in each time slot is directly related to the communication bandwidth, i.e., the larger the communication bandwidth is, the more symbols
the UAV can transmit within each time slot.
Following the above two aspects, we can make the number of channel uses in
each time slot large enough to be approximated as infinite by choosing appropriate communication bandwidth, flying altitude $H$, and the maximum flying speed $V_{\max}$. We note that this assumption (i.e., an infinite number of channel uses in each time slot) has also been widely used in the literature of UAV networks (e.g., \cite{JointWu2018,UnmanHWANG2018,Zhang2017Securing,Li2018Enabled}).
Thus, following \eqref{Hy_test} and \eqref{Like_Test}, $T_w[n]$ can be rewritten as
\begin{align}\label{T_w}
T_w[n]=
\begin{cases}
\sigma_w^2,&\Hnull,\\
\frac{\beta_0P_a[n]}{\|\mathbf{q}_a[n]-\mathbf{q}_w\|^2+H^2}+\sigma_w^2,&\Halt.
\end{cases}
\end{align}
In practice, a transceiver may not exactly know its AWGN power due to limited resources used to learn the dynamic environment. As such, following~\cite{goeckel2016covert,BiaoHe2017on}, in this work we consider noise uncertainty, i.e., $\sigma_w^2$ is a random variable with a known distribution. Specifically, we assume that $\sigma_{w,\mathrm{dB}}^2\in[\check{\sigma}_{\mathrm{dB}}^2-\varrho_{\mathrm{dB}},\check{\sigma}_{\mathrm{dB}}^2+\varrho_{\mathrm{dB}}]$, follows a uniform distribution within its value range in the $\mathrm{dB}$ domain, where $\sigma^2_{w,\mathrm{dB}}=10\log_{10}{(\sigma_w^2})$, $\check{\sigma}_{\mathrm{dB}}^2=10\log_{10}{(\check{\sigma}^2)}$, $\check{\sigma}^2$ denotes the nominal noise power, $\varrho_{\mathrm{dB}}=10\log_{10}{(\varrho)}$ is a parameter that measures the size of noise uncertainty, and $\varrho\geq 1$. Then, the distribution of $\sigma_w^2$ is given by
\begin{align}\label{D_sig}
f_{\sigma_w^2}(x)=
\begin{cases}
\frac{1}{2\ln{(\varrho)}x}, &\mathrm{if}~\frac{1}{\varrho}\check{\sigma}^2\leq x \leq \varrho\check{\sigma}^2,\\
0, &~\mathrm{otherwise}.
\end{cases}
\end{align}

In this work, we denote the false alarm rate and the miss detection rate at the $n$-th time slot by $P_F[n]\triangleq \mathrm{Pr}\{\Hoalt|\Hnull\}$ and $P_M[n]\triangleq \mathrm{Pr}\{\Honull|\Halt\}$, respectively, which can be written as
\begin{align}\label{PF}
P_F[n]=\mathrm{Pr}\{\sigma_w^2\geq P_{th}[n]\},
\end{align}
\begin{align}\label{PM}
P_M[n]=\mathrm{Pr}\Big\{\frac{\beta_0P_a[n]}{\|\mathbf{q}_a[n]-\mathbf{q}_w\|^2+H^2}+\sigma_w^2\leq P_{th}[n]\Big\},
\end{align}
respectively, $\forall n$. Then, the total error rate at the $n$-th time slot is given by
\begin{align}\label{Err_Pro}
\xi[n]=P_F[n]+P_M[n],~\forall n.
\end{align}

In covert communications, Willie wish to minimize the total error rate $\xi[n]$, $\forall n$.
In the following, we first derive the optimal detection threshold $P_{th}^*[n]$ from the perspective of Willie in order to achieve the minimum total error rate $\xi^*[n]$, and then we design the trajectory and the transmit power of UAV to ensure the average minimum total error rate being no less than a specific value, i.e., $\bar{\xi}^*[n]\geq 1-\rho_w$, where $\bar{\xi}^*[n]$ is achieved by averaging $\xi^*[n]$ over the priori distribution of Willie's location estimation error $\mathbf{e}_w$ and $\rho_w$ is an arbitrarily small value determining the required covertness.


%
%

%
\section{Detection Performance}

In this section, we first present Willie's false alarm rate and miss detection rate, based on which we obtain the optimal detection threshold $P_{th}^*[n]$ that minimizes the total error rate. Then, we derive the average total error rate at Willie from the perspective of UAV.

\subsection{Detection Performance at Willie}

According to \eqref{PF} and \eqref{PM}, at the $n$-th time slot, the false alarm and miss detection rates at Willie are given by
\begin{align}
P_F[n]&=
\begin{cases}\label{False}
1,&P_{th}[n]< \frac{\check{\sigma}^2}{\varrho},\\
\tau_1[n],& \frac{\check{\sigma}^2}{\varrho}\leq P_{th}[n] \leq {\varrho\check{\sigma}^2},\\
0,&P_{th}[n]> {\varrho\check{\sigma}^2},
\end{cases}\\
P_M[n]&=
\begin{cases}\label{Miss}
0,&P_{th}[n]< E_w[n]+\frac{\check{\sigma}^2}{\varrho},\\
\tau_2[n],&  E_w[n]+\frac{\check{\sigma}^2}{\varrho} \leq P_{th}[n]\leq E_w[n]+{\varrho\check{\sigma}^2},\\
1,&P_{th}[n]> E_w[n]+{\varrho\check{\sigma}^2},
\end{cases}
\end{align}
where $E_w[n]\triangleq \frac{\beta_0P_a[n]}{\|\mathbf{q}_a[n]-\mathbf{q}_w\|^2+H^2}$, and
\begin{align}\label{False_2}
\tau_1[n]\triangleq\int_{P_{th}[n]}^{\varrho\check{\sigma}^2}\frac{1}{2\ln{(\varrho)}x}dx=\frac{1}{2\ln{(\varrho)}}\ln{\left(\frac{\varrho\check{\sigma}^2}{P_{th}[n]}\right)},
\end{align}
\begin{align}\label{Miss_2}
\tau_2[n]\triangleq\int_{\frac{\check{\sigma}^2}{\varrho}}^{P_{th}[n]-E_w[n]}\frac{1}{2\ln{(\varrho)}x}dx=\frac{\ln{\left(\frac{\varrho(P_{th}[n]-E_w[n])}{\check{\sigma}^2}\right)}}{2\ln{(\varrho)}}.
\end{align}

In the following lemma, we derive the optimal detection threshold $P_{th}^*[n]$, $\forall n$, to minimize the total error rate $\xi[n]$ at Willie. We note that if $E_w[n]+\frac{\check{\sigma}^2}{\varrho}\geq \varrho\check{\sigma}^2$, Willie can set $P_{th}[n]=\varrho\check{\sigma}^2$ to ensure $\xi[n]=0$, which implies that Willie can detect any covert communication without any error. As such, in the following, we focus the case with $E_w[n]+\frac{\check{\sigma}^2}{\varrho}< \varrho\check{\sigma}^2$.

\begin{lemma}
The optimal detection threshold $P_{th}^*[n]$, which minimizes  $\xi[n]$ at Willie, is $E_w[n]+\frac{\check{\sigma}^2}{\varrho}$ and the corresponding minimum total error rate is given by
\begin{align}\label{DEP}
\xi^*[n]=\frac{1}{2\ln{(\varrho)}}\ln{\left(\frac{\varrho\check{\sigma}^2}{\frac{\beta_0P_a[n]}{\|\mathbf{q}_a[n]-\mathbf{q}_w\|^2+H^2}+\frac{\check{\sigma}^2}{\varrho}}\right)}.
\end{align}
\end{lemma}

\begin{IEEEproof}
The detailed proof is provided in Appendix A.
\end{IEEEproof}

\subsection{Detection Performance from UAV's Point of View}

In the last subsection, under the assumption that Willie can
obtain UAV's perfect location information, we derived Willie's
optimal detection threshold and the minimum total error rate. However, from UAV's point of view, it is difficult to obtain the exact location information of Willie, since Willie may be hidden somewhere on the ground, which motivate us to consider the location uncertainty at Willie from the perspective of UAV. As such, in the following, we derive the average total error rate over the Willie's uncertain locations as the measure on Willie's detection performance from UAV's point of view. To this end, we first introduce a random variable
\begin{align}\label{XN}
X[n]\triangleq\frac{1}{\varepsilon_w^2}\|\mathbf{q}_a[n]-\mathbf{\hat{q}}_w-\mathbf{e}_w\|^2,~\forall n,
\end{align}
which is the term $\|\mathbf{q}_a[n]-\mathbf{q}_w\|^2$ in \eqref{DEP} with location uncertainty at Willie, i.e., $\mathbf{q}_w = \mathbf{\hat{q}}_w+\mathbf{e}_w$. In the following lemma we derive the distribution of $X[n]$.

\begin{lemma}\label{lemma2}
The probability density function (pdf) of $X[n]$ is given by
\begin{align}\label{PDF_x}
f_{X[n]}(x)=
\begin{cases}
\frac{1}{2}e^{-\frac{1}{2}(x+\lambda[n])}\sum_{k=0}^{\infty}\frac{(\frac{\lambda[n]}{4})^kx^k}{k!k!},&x\geq 0,\\
0,&x<0,
\end{cases}
\end{align}
where $\lambda[n]$ is given by
\begin{align}\label{Lambda}
&\lambda[n]\triangleq \frac{1}{\varepsilon_w^2}\Big((x_a[n]-\hat{x}_w)^2+(y_a[n]-\hat{y}_w)^2\Big),~\forall n.
\end{align}
\end{lemma}
\begin{IEEEproof}
The detailed proof is provided in Appendix B.
\end{IEEEproof}

From Lemma~\ref{lemma2}, we know that  $X[n]$ follows noncentral chi-square distribution with 2 degrees of freedom~\cite{FunKay1998}. The expectation and variance of the $X[n]$ are given by $\lambda[n]+2$ and $4\lambda[n]+4$, respectively, where $\lambda[n]$ is the noncentral parameter.

Following the total error rate $\xi^*[n]$ given in \eqref{DEP} and Lemma~2, the average minimum total error rate over Willie's location uncertainty is given by
\begin{align}\label{Ave_1}
&\bar{\xi}^*[n]=\int_0^{\infty}\frac{1}{2\ln{(\varrho)}}\ln{\left(\frac{\varrho\check{\sigma}^2}{\frac{\frac{\beta_0}{\varepsilon_w^2}P_a[n]}{x+\frac{H^2}{\varepsilon_w^2}}+\frac{\check{\sigma}^2}{\varrho}}\right)}f_{X[n]}(x)dx\nonumber\\
&=1-\int_0^{\infty}\frac{1}{2\ln{(\varrho)}}\ln{\left(1+\frac{\beta_1P_a[n]}{x+\frac{H^2}{\varepsilon_w^2}}\right)}f_{X[n]}(x)dx,\notag\\
&=1-\mathbb{E}_{X[n]}\left[\frac{1}{2\ln{(\varrho)}}\ln{\left(1+\frac{\beta_1P_a[n]}{X[n]+\frac{H^2}{\varepsilon_w^2}}\right)}\right]\nonumber\\
&=1-\frac{1}{2\ln{(\varrho)}}\mathbb{E}_{X[n]}\left[g(X[n])\right],
\end{align}
where $\beta_1\triangleq\frac{\beta_0\varrho}{\varepsilon_w^2\check{\sigma}^2}$ and $g(X[n])\triangleq\ln{\Big(1+\frac{\beta_1P_a[n]}{X[n]+\frac{H^2}{\varepsilon_w^2}}\Big)}$.
We note that $g(X[n])$ is a convex function with respect to $X[n]$.

Since the pdf of $X[n]$ is very complex (as shown in Lemma~2) and $X[n]$ is in the denominator of the logarithmic function, an exact analytical expression of \eqref{Ave_1} is mathematically intractable.
In the context of covert communication, it is required $\bar{\xi}^\ast[n]\geq 1-\rho_w$, $\forall n$, for an arbitrarily small $\rho_w$, in order to guarantee the covertness of UAV's transmission. Considering the intractable expression of $\bar{\xi}^\ast[n]$ and the required covertness constraint $\bar{\xi}^\ast[n]\geq 1-\rho_w$, in the following we derive a lower bound on $\bar{\xi}^\ast[n]$. To this end, we first present the following lemma to derive an upper bound on $\mathbb{E}_{X[n]}\left[g(X[n])\right]$.

\begin{lemma}\label{lemma3}
If $f(x)$ is a convex function with respect to the variable $x \in [a, b]$ with $\mu_x$ and $\sigma^2_x$ as the expectation and variance of $x$, respectively, we have
\begin{align}\label{lemma2_1}
\mathbb{E}_x[f(x)]\leq p_1f(a)+p_2f(\mu_x)+p_3f(b),
\end{align}
where $p_1\triangleq\frac{\theta}{2(\mu_x-a)}$, $p_2\triangleq 1-\frac{\theta}{2(\mu_x-a)}-\frac{\theta}{2(b-\mu_x)}$, $p_3\triangleq\frac{\theta}{2(b-\mu_x)}$, and $\theta\triangleq\mathbb{E}(|x-\mu_x|)$ denotes the mean absolute deviation of $x$. As $b\rightarrow\infty$, \eqref{lemma2_1} can be written as
\begin{align}\label{lemma2_2}
\mathbb{E}_x[f(x)]\leq f(\mu_x)\!+\!\frac{\theta}{2}\left(\lim_{t\rightarrow\infty}\frac{f(t)}{t}+\frac{f(a)-f(\mu_x)}{\mu_x-a}\right).
\end{align}
When $x\sim\mathbb{N}(\mu_x,\sigma_x^2)$, we have $\theta=\sqrt{\frac{2}{\pi}}\sigma_x$.
\end{lemma}

\begin{IEEEproof}
The detailed proof is presented in~\cite{Ben1972More,Ben1985Approximation}.
\end{IEEEproof}

To use Lemma~\ref{lemma3} for determining a lower bound on the minimum detection error rate $\bar{\xi}^\ast[n]$, we next show that the random variable $X[n]$ of interest can be precisely approximated as a Gaussian random variable in the considered scenario. To this end, we first present the following lemma.

\begin{lemma}\label{lemma4}
For a noncentral chi-square random variable $V=\sum_{i=1}^kv_i^2$ with $\eta$ as the noncentral parameter, its pdf can be precisely approximated as the Gaussian distribution $\mathbb{N}(k+\eta, 2k+4\eta)$, even when the degree of freedom $k$ is relatively small.
\end{lemma}

\begin{IEEEproof}
The detailed proof is provided in Appendix C.
\end{IEEEproof}

Following Lemma~\ref{lemma2}, we know that $\lambda[n]$ given \eqref{Lambda} is the noncentral parameter of $X[n]$. We note that the value of $\lambda[n]$ is mainly determined by the distance from UAV to Willie's estimated location, which is normally large in covert communications to ensure the required covertness. As such, we can conclude that $\lambda[n]$ is usually very large. Then, following Lemma~\ref{lemma4}, the pdf of $X[n]$ given in Lemma~\ref{lemma2} can be approximated as $\mathbb{N}(2+\lambda[n],4+4\lambda[n])$. Then, following Lemma~\ref{lemma3} and Lemma~\ref{lemma4}, we determine a lower bound on the minimum detection error rate $\bar{\xi}^\ast[n]$ in the following proposition.

\begin{proposition}\label{proposition1}
A lower bound on the average minimum total error rate $\bar{\xi}^*[n]$ is given by
\begin{align}\label{lower}
\check{\xi}^*[n] \leq \bar{\xi}^*[n],
\end{align}
where
\begin{align}\label{def_xi}
&\check{\xi}^*[n] =  1-\frac{1}{2\ln{(\varrho)}}g^u(\lambda[n],P_a[n]),
\end{align}
\begin{align}\label{g_u}
&g^u(\lambda[n],P_a[n])\!=\! \ln\left(1\!+\!\frac{\beta_1P_a[n]}{\lambda[n]\!+\!2\!+\!\frac{H^2}{\varepsilon_w^2}}\right)\!+\!\frac{\sqrt{2}}{\sqrt{\pi}\sqrt{\lambda[n]+1}}\notag\\
&\times \left(\ln\left(1\!+\!\frac{\beta_1\varepsilon_w^2 P_a[n]}{H^2}\right)\!-\!\ln\left(1\!+\!\frac{\beta_1P_a[n]}{\lambda[n]\!+\!2\!+\!\frac{H^2}{\varepsilon_w^2}}\right)\right).
\end{align}
\end{proposition}

\begin{IEEEproof}
The detailed proof is provided in Appendix D.
\end{IEEEproof}

Following Proposition~\ref{proposition1} and recalling that the original covertness constraint is $\bar{\xi}^*[n] \geq 1 - \rho_w$, in the following sections we use $\check{\xi}^*[n] \geq 1 - \rho_w$ as our covertness constraint in our considered optimization problems.

\section{Joint Optimization of UAV's Trajectory and Transmit Power for Covert Communications}

In this section, we aim to design the trajectory and transmit power of UAV to maximize the transmission rate from UAV to Bob subject to a transmission outage constraint and the aforementioned covertness constraint.

\subsection{ Optimization Problem Formulation}

Considering location uncertainty, the transmission from UAV to Bob may occur outage. In practice, many wireless communication systems can tolerate a certain transmission outage probability that does not significantly affect the desired quality of service (QoS). As such, in this work we require this transmission outage probability being greater than or equal to a specific value, which serves as a constraint in the optimization of the UAV's trajectory and transmit power. Considering the UAV's mobility constraint, power constraint, and the covertness constraint, the optimization problem at the UAV is formulated as
\begin{subequations}\label{PF1}
\begin{align}
&(\mathrm{P}1):~\max_{\mathbf{Q}_a,\mathbf{P}_a,\mathbf{R}_b}~\frac{1}{N}\sum_{n=1}^N R_b[n]\label{PF1a}\\
&\mathrm{s.t.}~\mathrm{Pr}\left\{\log_2\Big(1+\frac{P_a[n]\gamma_0}{\|\mathbf{q}_a[n]-\mathbf{\hat{q}}_b-\mathbf{e}_b\|^2+H^2}\Big)\geq R_b[n]\right\}\nonumber\\
&~~~~~~~~~~~~~~~~~~~~~~~~~~~~~~~~~~~~~~~~~~~~~~\geq 1-\rho_b,\forall n,\label{PF1b}\\
&~~~~~\check{\xi}^*[n]\geq 1-\rho_w,\label{PF1c}\\
&~~~~~\eqref{Mob_C},\label{PF1d}\\
&~~~~~\eqref{Pow_C},\label{PF1e}
\end{align}
\end{subequations}
where $\mathbf{Q}_a\triangleq\{\mathbf{q}_a[n],\forall n\}$, $\mathbf{P}_a\triangleq\{P_a[n],\forall n\}$, $\mathbf{R}_b\triangleq\{R_b[n],\forall n\}$, and $\rho_b$ is the maximum allowable transmission outage probability from UAV to Bob. We note that, $\gamma_0$ in (26b) is defined as $\frac{\beta_0}{\sigma_b^2}$, where $\sigma_b^2$ is the power of the AWGN at Bob. The objective function given in \eqref{PF1a} is the average transmission rate over the $N$ time slots of interest and maximizing it means to maximize the total amount of the transmitted information from UAV to Bob over the $N$ time slots. The transmission outage probability constraint \eqref{PF1b} ensures a certain required QoS between UAV and Bob. In addition, the constraint \eqref{PF1c} is to ensure the covertness of the transmission from UAV to Bob. Furthermore, the constraint \eqref{PF1d} detailed in \eqref{Mob_C} is the mobility constraint on the UAV, while the constraint \eqref{PF1e} detailed in \eqref{Pow_C} serves as the transmit power constraint at the UAV.

We note that the objective function is a linear function of $R_b[n]$, $\forall n$, while the mobility constraint \eqref{PF1d} and the power constraint \eqref{PF1e} are convex with respect to {{$\mathbf{q}_a[n]$ and $P_a[n]$, respectively}}. However, the transmission outage probability constraint \eqref{PF1b} and covertness constraint \eqref{PF1c} are non-convex. As such, the optimization problem (P1) is difficult to tackle directly. Therefore, in the following we develop an algorithm to solve it step by step.

\subsection{On the Constraints of the Optimization Problem (P1)}

The main challenge to solve the optimization problem (P1) arises from the intractable expression for the outage probability involved in the constraint \eqref{PF1b} and the complex expression for the minimum average total error rate involved in the constraint \eqref{PF1c}. In the following, we aim to tackle the non-convex constraints \eqref{PF1b} and \eqref{PF1c}, respectively, in order to facilitate solving (P1).

\subsubsection{Transmission Outage Probability Constraint \eqref{PF1b}}

We note that it is difficult to directly derive the analytical expression for the outage probability in \eqref{PF1b}.
Fortunately, we find that the Bernstein-type inequality (BTI) \cite{KY2014Outage} can transform this constraint into a deterministic form. To proceed, we present the BTI in the following lemma.

\begin{lemma}\label{lemma5}
If $\mathbf{A}\in\mathbb{C}^{N\times N}$ is a Hermitian matrix, $\mathbf{x}\sim\mathcal{CN}(0,\mathbf{I})$, $\mathbf{w}\in\mathbb{C}^{N\times 1}$, $\rho\in(0,1]$, and $c\in \mathbb{R}$ is independent of $\mathbf{x}$, the following implication holds:
\begin{align}\label{BTI}
\mathrm{Pr}\Big\{\mathbf{x}^H\mathbf{A}\mathbf{x}+2\mathfrak{R}\big\{\mathbf{x}^H\mathbf{w}\big\}+c\geq 0\Big\}\geq 1-\rho.\\
\Leftarrow
\begin{cases}
\mathrm{Tr}(\mathbf{A})-\sqrt{-2\ln{\rho}}z+\ln{\rho}y+c\geq 0,\\
\begin{Vmatrix}
\mathrm{vec}(\mathbf{A})\\\sqrt{2}\mathbf{w}
\end{Vmatrix}\leq z\\
y\mathbf{I}_N+\mathbf{A}\succeq \mathbf{0},
\end{cases}
\end{align}
where $y$ and $z$ are slack variables.
\end{lemma}
\begin{IEEEproof}
The detailed proof can be found in \cite{KY2014Outage}.
\end{IEEEproof}

In order to apply the BTI, we first rewrite the outage constraint \eqref{PF1b} as
\begin{align}\label{Pout1}
\mathrm{Pr}\Big\{\|\mathbf{q}_a[n]\!-\!\mathbf{\hat{q}}_b\!-\!\mathbf{e}_b\|^2\leq \frac{P_a[n]\gamma_0}{2^{R_b[n]}-1}\!-\!H^2\Big\}\geq 1-\rho_b,~\forall n.
\end{align}
Recall that $\mathbf{e}_b\sim\mathbb{N}(\mathbf{0},\varepsilon_b^2\mathbf{I}_2)$. As such, we have $\mathbf{e}_b=\varepsilon_b\mathbf{u}$, where $\mathbf{u}\sim\mathbb{N}(\mathbf{0},\mathbf{I}_2)$. Consequently, \eqref{Pout1} can be rewritten as
\begin{align}\label{Pout2}
&\mathrm{Pr}\Big\{-\varepsilon_b^2\mathbf{u}^T\mathbf{u}+2\varepsilon_b\mathbf{u}^T(\mathbf{q}_a[n]-\mathbf{\hat{q}}_b)-\mathbf{q}_a^T[n]\mathbf{q}_a[n]\\
&+2\mathbf{q}_a^T[n]\mathbf{\hat{q}}_b-\mathbf{\hat{q}}_b^T\mathbf{\hat{q}}_b+\frac{P_a[n]\gamma_0}{2^{R_b[n]}-1}-H^2\geq 0\Big\}\geq 1\!-\!\rho_b,\forall n.\nonumber
\end{align}
We observe that, the constraint \eqref{Pout2} is in a similar form as \eqref{BTI}. As such, following Lemma~\ref{lemma5} and \eqref{Pout2}, we have
\begin{subequations}\label{Pout3}
\begin{align}
&\mathrm{Tr}\left(-\varepsilon_b^2\mathbf{I}_2\right)-\sqrt{-2\ln{\rho_b}}z_1[n]+\ln{\rho_b}y_1[n]-\mathbf{q}_a^T[n]\mathbf{q}_a[n]\nonumber\\
&~~+2\mathbf{q}_a^T[n]\mathbf{\hat{q}}_b-\mathbf{\hat{q}}_b^T\mathbf{\hat{q}}_b+\frac{P_a[n]\gamma_0}{2^{R_b[n]}-1}-H^2\geq 0,~\forall n,\label{Pout3a}\\
&\begin{Vmatrix}
\mathrm{vec}\left(-\varepsilon_b^2\mathbf{I}_2\right)\\ \sqrt{2}\varepsilon_b(\mathbf{q}_a[n]-\mathbf{\hat{q}}_b)
\end{Vmatrix}\leq z_1[n],~\forall n,\label{Pout3b}\\
&y_1[n]\mathbf{I}_2-\varepsilon_b^2\mathbf{I}_2\succeq \mathbf{0},~y_1[n]\geq 0,~z_1[n]\geq 0,~\forall n,\label{Pout3c}
\end{align}
\end{subequations}
where $y_1[n]$ and $z_1[n]$ are introduced slack variables. We note that, \eqref{Pout3c} can be further simplified as
\begin{align}\label{Pout4}
y_1[n]\geq \varepsilon_b^2,~z_1[n]\geq 0,~\forall n.
\end{align}

Replacing the outage constraint \eqref{PF1b} in (P1) with \eqref{Pout3a}, \eqref{Pout3b} and \eqref{Pout4}, the optimization problem (P1) can be rewritten as
\begin{align}\label{PF1_1}
&(\mathrm{P}1.1):~\max_{\mathbf{Q}_a,\mathbf{P}_a,\mathbf{R}_b,\mathbf{Y}_1,\mathbf{Z}_1}~\frac{1}{N}\sum_{n=1}^N R_b[n]\nonumber\\
&\mathrm{s.t.}~\eqref{Pout3a},~\eqref{Pout3b},~\eqref{Pout4},~\eqref{PF1c},~\eqref{Mob_C},~\eqref{Pow_C},
\end{align}
where $\mathbf{Y}_1\triangleq\{y_1[n],\forall n\}$ and $\mathbf{Z}_1\triangleq\{z_1[n],\forall n\}$.
We note that all the constraints in (P1.1)  are in exact analytical expressions.
We also note that \eqref{Pout3b} is a second-order cones (SOC) constraint, and \eqref{Pout4} is a linear constraint, both of which are convex. However, \eqref{Pout3a} is still non-convex due to the non-concavity of the term $\frac{P_a[n]\gamma_0}{2^{R_b[n]}-1}$. In addition, the total error rate constraint \eqref{PF1c} is also non-convex. In the following, we apply the first-order restrictive approximation to transform (P1.1) into a convex optimization problem and then using the SCA algorithm to solve it. For now, we focus on transforming \eqref{Pout3a} into a convex constraint. We observe from \eqref{Pout3a} that the optimization variables $P_a[n]$ and $R_b[n]$  are coupled in the term $\frac{P_a[n]\gamma_0}{2^{R_b[n]}-1}$. To proceed with the transformation, we first present the following lemma.
\begin{lemma}\label{lemma6}
If $R_b[n]>0$, $\forall n$, then the following inequality must hold
\begin{align}\label{Pout5}
&\frac{\gamma_0 P_a[n]}{2^{R_b[n]}-1}\geq g_1\left(P_a[n],R_b[n],\tilde{P}_a[n],\tilde{R}_b[n]\right),
\end{align}
where
\begin{align}\label{Pout5_2}
&g_1\left(P_a[n],R_b[n],\tilde{P}_a[n],\tilde{R}_b[n]\right)= \frac{\tilde{P}_a[n]\gamma_0}{2^{\tilde{R}_b[n]}-1}-\frac{(\tilde{P}_a[n])^{2}\gamma_0}{(2^{\tilde{R}_b[n]}-1)}\notag\\
&\times \left(\frac{1}{P_a[n]}-\frac{1}{\tilde{P}_a[n]}\right)-
\frac{\tilde{P}_a[n]\gamma_0(2^{R_b[n]}-2^{\tilde{R}_b[n]})}{(2^{\tilde{R}_b[n]}-1)^2},
\end{align}
while $\tilde{P}_a[n]$ and $\tilde{R}_b[n]$, $\forall n$, are given feasible points.
\end{lemma}

\begin{IEEEproof}
The detailed proof is provided in Appendix E.
\end{IEEEproof}

We note that, for given feasible points $(\tilde{P}_a[n],\tilde{R}_b[n])$, $\forall n$, $g_1\left(P_a[n],R_b[n],\tilde{P}_a[n],\tilde{R}_b[n]\right)$ is a concave function with respect to the optimization variables $P_a[n]$ and $R_b[n]$. As such, the non-convex constraint \eqref{Pout3a} can be rewritten as
\begin{align}\label{Pout6}
&\mathrm{Tr}(-\varepsilon_b^2\mathbf{I}_2)\!\!-\!\!\sqrt{\!-\!2\ln{\rho_b}}z_1[n]\!+\!\ln{\rho_b}y_1[n]\!-\!\mathbf{q}_a^T[n]\mathbf{q}_a[n]\!-\!\mathbf{\hat{q}}_b^T\mathbf{\hat{q}}_b\nonumber\\
&\!+\!2\mathbf{q}_a^T[n]\mathbf{\hat{q}}_b\!+\!
g_1\left(P_a[n],R_b[n],\tilde{P}_a[n],\tilde{R}_b[n]\right)\!-\!H^2\!\geq\! 0,
\end{align}
$\forall n$, which is convex. So far, we have transformed the non-convex constraint \eqref{PF1b} into a convex form, which is detailed in \eqref{Pout3b}, \eqref{Pout4}, and \eqref{Pout6}.

\newcounter{mytempeqncnt2}
\begin{figure*}[tp]
\normalsize
\setcounter{mytempeqncnt2}{\value{equation}}
\setcounter{equation}{40}
\begin{align}\label{Det5}
&\sqrt{\left(\frac{1}{\varepsilon_w^2}\|\mathbf{q}_a[n]-\mathbf{\hat{q}}_w\|^2+1\right)u_1[n]}\leq\sqrt{\left(\frac{1}{\varepsilon_w^2}\|\mathbf{\tilde{q}}_a[n]-\mathbf{\hat{q}}_w\|^2+1\right)\tilde{u}_1[n]}+
\frac{1}{2}\sqrt{\frac{\frac{1}{\varepsilon_w^2}\|\mathbf{\tilde{q}}_a[n]-\mathbf{\hat{q}}_w\|^2+1}{\tilde{u}_1[n]}}(u_1[n]-\tilde{u}_1[n])+\nonumber\\
&\frac{1}{2}\sqrt{\frac{\tilde{u}_1[n]}{\frac{1}{\varepsilon_w^2}\|\mathbf{\tilde{q}}_a[n]-\mathbf{\hat{q}}_w\|^2+1}}
\left(\frac{1}{\varepsilon_w^2}\|\mathbf{q}_a[n]-\mathbf{\hat{q}}_w\|^2-\frac{1}{\varepsilon_w^2}\|\mathbf{\tilde{q}}_a[n]-\mathbf{\hat{q}}_w\|^2\right)\triangleq g_2(\mathbf{q}_a[n],u_1[n],\mathbf{\tilde{q}}_a[n],\tilde{u}_1[n]),
\end{align}
\setcounter{equation}{45}
\begin{align}\label{Det9}
&\ln\left(\frac{1}{\varepsilon_w^2}\|\mathbf{q}_a[n]-\mathbf{\hat{q}}_w\|^2+2+\frac{H^2}{\varepsilon_w^2}+\beta_1P_a[n]\right)\leq \ln\left(\frac{1}{\varepsilon_w^2}\|\mathbf{\tilde{q}}_a[n]-\mathbf{\hat{q}}_w\|^2+2+\frac{H^2}{\varepsilon_w^2}+\beta_1\tilde{P}_a[n]\right)+\nonumber\\
&\frac{\left(\frac{1}{\varepsilon_w^2}\|\mathbf{q}_a[n]-\mathbf{\hat{q}}_w\|^2-\frac{1}{\varepsilon_w^2}\|\mathbf{\tilde{q}}_a[n]-\mathbf{\hat{q}}_w\|^2\right)+\beta_1(P_a[n]-\tilde{P}_a[n])}{\frac{1}{\varepsilon_w^2}\|\mathbf{\tilde{q}}_a[n]-\mathbf{\hat{q}}_w\|^2+2+\frac{H^2}{\varepsilon_w^2}+\beta_1\tilde{P}_a[n]}\triangleq g_4(\mathbf{q}_a[n],P_a[n],\mathbf{\tilde{q}}_a[n],\tilde{P}_a[n]),~\forall n.
\end{align}
\setcounter{equation}{\value{mytempeqncnt2}}
\hrulefill
\vspace*{4pt}
\end{figure*}

\subsubsection{Covertness Constraint \eqref{PF1c}}

Following {{\eqref{def_xi}}}, we can rewrite the constraint \eqref{PF1c} as
\begin{align}\label{Det1}
g^u(\lambda[n],P_a[n])\leq 2\ln{(\varrho)}\rho_w,~\forall n,
\end{align}
where $g^u(\lambda[n],P_a[n])$ is defined in {{\eqref{g_u}}}. We note that the expression of $g^u(\lambda[n],P_a[n])$ is very complex and the optimization variables $P_a[n]$ and $\mathbf{q}_a[n]$, $\forall n$, are coupled, which leads to the fact that \eqref{PF1c} is hard to tackle. To proceed, we first multiply $\sqrt{\frac{\pi}{2}}\sqrt{\lambda[n]+1}$ on both sides of the constraint \eqref{Det1}, yielding
\begin{align}\label{Det2}
&\left(\sqrt{\frac{\pi}{2}}\sqrt{\lambda[n]+1}-1\right)\ln\left(1+\frac{\beta_1P_a[n]}{\lambda[n]+2+\frac{H^2}{\varepsilon_w^2}}\right)+\nonumber\\
&\ln\left(1+\frac{\beta_1\varepsilon_w^2 P_a[n]}{H^2}\right)\leq \sqrt{2\pi}\ln{(\varrho)}\rho_w\sqrt{\lambda[n]+1},
\end{align}
 $\forall n$. {{We note that $\left(\sqrt{\frac{\pi}{2}}\sqrt{\lambda[n]+1}-1\right)>0$ must hold, since $\lambda[n]$ defined in \eqref{Lambda} must be nonnegative.}} Introducing the slack variables $u_1[n]$ and $u_2[n]$, $\forall n$, \eqref{Det2} can be rewritten as
\begin{subequations}\label{Det3}
\begin{align}
&\left(\sqrt{\frac{\pi}{2}}\sqrt{\lambda[n]+1}-1\right)\sqrt{u_1[n]}+\nonumber\\
&~\ln\left(1+\frac{\beta_1\varepsilon_w^2 P_a[n]}{H^2}\right)\leq \sqrt{2\pi}\ln{(\varrho)}\rho_w\sqrt{u_2[n]},~\forall n,\label{Det3a}\\
&\ln\left(1+\frac{\beta_1P_a[n]}{\lambda[n]+2+\frac{H^2}{\varepsilon_w^2}}\right)\leq \sqrt{u_1[n]},~\forall n,\label{Det3b}\\
&\lambda[n]+1\geq u_2[n],~\forall n.\label{Det3c}
\end{align}
\end{subequations}

We note that the equalities in \eqref{Det3a}, \eqref{Det3b}, and \eqref{Det3c} must hold for the optimal solution. Otherwise, the objective function can be further improved by adjusting the slack variables.
Although \eqref{Det3a}, \eqref{Det3b}, and \eqref{Det3c} are still non-convex, they are all in tractable forms compared to \eqref{Det2}. In the following, we focus on transforming \eqref{Det3a}, \eqref{Det3b}, and \eqref{Det3c} into convex constraints.
We first rewrite the constraint \eqref{Det3a} as
\begin{align}\label{Det4}
&\sqrt{\frac{\pi}{2}}\sqrt{
\left(\frac{1}{\varepsilon_w^2}\|\mathbf{q}_a[n]-\mathbf{\hat{q}}_w\|^2+1\right)u_1[n]}-\sqrt{u_1[n]}+\nonumber\\
&~~\ln\left(1+\frac{\beta_1\varepsilon_w^2 P_a[n]}{H^2}\right)\leq \sqrt{2\pi}\ln{(\varrho)}\rho_w\sqrt{u_2[n]},~\forall n.
\end{align}
We note that the first term and the third term in the left hand side (LHS) of \eqref{Det4} are non-convex. As such, in order to transform \eqref{Det4} into a convex constraint, we first present the following lemma.

\begin{lemma}
For $u_1[n]>0$, $\forall n$, the inequality \eqref{Det5} must hold, which is shown at the top of the next page,
where $\mathbf{\tilde{q}}_a[n]$ and $\tilde{u}_1[n]$, $\forall n$, are given feasible points.
\end{lemma}

\begin{IEEEproof}
The detailed proof is provided in Appendix F.
\end{IEEEproof}
We recall that $\frac{1}{\varepsilon_w^2}\|\mathbf{q}_a[n]-\mathbf{\hat{q}}_w\|^2$ is a convex function with respect to $\mathbf{q}_a[n]$. For given feasible points $(\mathbf{\tilde{q}}_a[n],\tilde{u}_1[n])$, $\forall n$,
\eqref{Det5} is a linear function with respect to $u_1[n]$ and it is a convex function with respect to $\mathbf{q}_a[n]$. {{As such, \eqref{Det5} is jointly convex with respect to $\mathbf{q}_a[n]$ and $u_1[n]$.}} In addition, we observe that $\ln\big(1+\frac{\beta_1\varepsilon_w^2 P_a[n]}{H^2}\big)$ in \eqref{Det4} is a concave function with respect to $P_a[n]$. {{We note that the first-order approximation of any concave function is its upper bound~\cite{Boyd}.}}
As such, for given feasible points $\tilde{P}_a[n]$, $\forall n$, we have
\setcounter{equation}{41}
\begin{align}\label{Det6}
&\ln\left(1+\frac{\beta_1\varepsilon_w^2P_a[n]}{H^2}\right)\leq g_3(P_a[n],\tilde{P}_a[n]),~\forall n,
\end{align}
where
\begin{align}\label{Det6_2}
&g_3(P_a[n],\tilde{P}_a[n]) =  \notag \\
&\ln\left(1+\frac{\beta_1\varepsilon_w^2 \tilde{P}_a[n]}{H^2}\right)+\frac{\beta_1\varepsilon_w^2(P_a[n]-\tilde{P}_a[n])}{H^2+\beta_1\varepsilon_w^2\tilde{P}_a[n]},~\forall n.
\end{align}
We note that, for given feasible points $\tilde{P}_a[n]$, $\forall n$,  $g_3(P_a[n],\tilde{P}_a[n])$ is a linear function with respect to the transmit power $P_a[n]$. As such, $g_3(P_a[n],\tilde{P}_a[n])$ is a convex function with respect to {{$P_a[n]$.}} According to \eqref{Det5} and \eqref{Det6}, the restrictive approximation of \eqref{Det4} is given by
\begin{align}\label{Det7}
&\sqrt{\frac{\pi}{2}}g_2(\mathbf{q}_a[n],u_1[n],\mathbf{\tilde{q}}_a[n],\tilde{u}_1[n])-\sqrt{u_1[n]}+\nonumber\\
&~~~~~~~g_3(P_a[n],\tilde{P}_a[n])\leq \sqrt{2\pi}\ln{(\varrho)}\rho_w\sqrt{u_2[n]},~\forall n.
\end{align}
We note that the LHS of \eqref{Det7} is convex with respect to $\mathbf{q}_a[n]$ and $u_1[n]$. Since $\sqrt{u_2[n]}$ is a concave function with respect to $u_2[n]$, the right hand side (RHS) of \eqref{Det7} is concave with respect to {{$u_2[n]$.}} As such, we now have transformed the non-convex constraint \eqref{Det3a}  into a convex constraint given in \eqref{Det7}.

In order to transform the constraint \eqref{Det3b} into a convex one, we first rewrite it as
\begin{align}\label{Det8}
&\ln\left(\frac{1}{\varepsilon_w^2}\|\mathbf{q}_a[n]-\mathbf{\hat{q}}_w\|^2+2+\frac{H^2}{\varepsilon_w^2}+\beta_1P_a[n]\right)-\nonumber\\
&~~~~~~~~~~~~~~~\ln\left(u_2[n]+1+\frac{H^2}{\varepsilon_w^2}\right)\leq \sqrt{u_1[n]},~\forall n.
\end{align}
We observe that the first term in the LHS of \eqref{Det8} is jointly concave with respect to $\frac{1}{\varepsilon_w^2}\|\mathbf{q}_a[n]-\mathbf{\hat{q}}_w\|^2$ and $P_a[n]$, and the second term in the LHS of \eqref{Det8} is convex with respect to $u_2[n]$.
This special form allows us to apply the first-order
approximation to transform \eqref{Det8} into a convex constraint. Similar to \eqref{Det6}, for given feasible points $(\mathbf{\tilde{q}}_a[n],\tilde{P}_a[n])$, $\forall n$, the first-order restrictive approximation of the first term in the LHS of \eqref{Det8} is given by \eqref{Det9}, which is shown at the top of this page. We observe that, for given feasible points $(\mathbf{q}_a[n],\tilde{P}_a[n])$, $\forall n$, \eqref{Det9} is a convex function with respect to $\mathbf{q}_a[n]$ and $P_a[n]$. As such, the first-order restrictive approximation of \eqref{Det8} is given by
\setcounter{equation}{46}
\begin{align}\label{Det10}
&g_4(\mathbf{q}_a[n],P_a[n],\mathbf{\tilde{q}}_a[n],\tilde{P}_a[n])-\nonumber\\
&~~~~~~~~~~~~~~~\ln\left(u_2[n]+1+\frac{H^2}{\varepsilon_w^2}\right)\leq \sqrt{u_1[n]},~\forall n,
\end{align}
{{which is jointly convex with respect to $\mathbf{q}_a[n]$, $P_a[n]$, and $u_1[n]$.}}

Finally, we tackle the constraint \eqref{Det3c} in order to converge it into a convex constraint. We note that \eqref{Det3c} is non-convex due to the super-level set of the convex quadratic function  $\frac{1}{\varepsilon_w^2}\|\mathbf{q}_a[n]-\mathbf{\hat{q}}_w\|^2$.
We recall that any convex function is lower bounded by its first-order approximation at any feasible point~\cite{Boyd}. As such, for given feasible points $\mathbf{\tilde{q}}_a[n]$, $\forall n$, the first-order lower bound of $\frac{1}{\varepsilon_w^2}\|\mathbf{q}_a[n]-\mathbf{\hat{q}}_w\|^2$ is given by
\begin{align}\label{Det11}
&\frac{1}{\varepsilon_w^2}\|\mathbf{q}_a[n]-\mathbf{\hat{q}}_w\|^2\geq \frac{2}{\varepsilon_w^2}(\mathbf{\tilde{q}}_a[n]-\mathbf{\hat{q}}_w)^T(\mathbf{q}_a[n]-\mathbf{\tilde{q}}_a[n])+ \nonumber\\
&~~~~~~~~~~~~~\frac{1}{\varepsilon_w^2}\|\mathbf{\tilde{q}}_a[n]-\mathbf{\hat{q}}_w\|^2\triangleq g_5(\mathbf{q}_a[n],\mathbf{\tilde{q}}_a[n]),~\forall n.
\end{align}
We note that \eqref{Det11} is a linear function with respect to the optimization variable $\mathbf{q}_a[n]$. As such, for given feasible points $\mathbf{\tilde{q}}_a[n]$, $\forall n$,
the first-order restrictive approximation of \eqref{Det3c} is given by
\begin{align}\label{Det12}
g_5(\mathbf{q}_a[n],\mathbf{\tilde{q}}_a[n])+1\geq u_2[n],~\forall n,
\end{align}
which is convex.

So far, we have transformed the non-convex covertness constraint \eqref{PF1c} into a convex form, which is detailed by \eqref{Det7}, \eqref{Det10}, and \eqref{Det12}. Following the above transformation, the optimization problem (P1.1) can be reformulated as
\begin{align}\label{PF1_2}
&(\mathrm{P}1.2):~\max_{\mathbf{Q}_a,\mathbf{P}_a,\mathbf{R}_b,\mathbf{Y}_1,\mathbf{Z}_1,\mathbf{U}_1,\mathbf{U}_2}~\frac{1}{N}\sum_{n=1}^N R_b[n]\nonumber\\
&\mathrm{s.t.}~\eqref{Pout3b},~\eqref{Pout4},~\eqref{Pout6},~\eqref{Det7}, ~\eqref{Det10},~\eqref{Det12},~\eqref{Mob_C},~\eqref{Pow_C},
\end{align}
where $\mathbf{U}_1\triangleq\{u_1[n],\forall n\}$ and $\mathbf{U}_2\triangleq\{u_2[n],\forall n\}$. In the following subsection, we present the algorithm used to solve the optimization problem $(\mathrm{P}1.2)$.

\subsection{Overall Algorithm}

The optimization problem (P1.2) is with a linear objective function and a convex constraint set. As such, it is a convex optimization problem.
For given feasible points $(\mathbf{\tilde{q}}_a[n],\tilde{P}_a[n],\tilde{R}_b[n],\tilde{u}_1[n])$, $\forall n$, the optimization problem (P1.2) can be efficiently solved by convex optimization solver such as CVX\cite{Boyd}.
We note that the constraint set of (P1.2) is more stricter than the constraint set of the original optimization problem (P1.1). This leads to that the solution to the optimization problem (P1.2) is also feasible to (P1.1). We note that the two optimization problems are equivalent at the given feasible points
$(\mathbf{\tilde{q}}_a[n],\tilde{P}_a[n],\tilde{R}_b[n],\tilde{u}_1[n])$, $\forall n$.
We also note that the objective value of (P1.1) can be further improved by successively solving the optimization problem (P1.2).
According to the principle of SCA, at each iteration, the current solution to the optimization problem (P1.2) gradually approximates the solution to the optimization problem (P1.1) convergely.
The algorithm used to solve the optimization problem (P1.1) is detailed in  Algorithm~\ref{alg1}.
\begin{algorithm}[t]
\caption{SCA Algorithm for Solving Problem (P1.1)}\label{alg1}
\begin{algorithmic}[1]
\STATE Initialize: Given feasible $(\mathbf{\tilde{q}}_a^0[n],\tilde{P}_a^0[n],\tilde{R}_b^0[n],\tilde{u}_1^0[n])$, $\forall n$; $i=0$.
\REPEAT
\STATE {Given a feasible $(\mathbf{\tilde{q}}_a^i[n],\tilde{P}_a^i[n],\tilde{R}_b^i[n],\tilde{u}_1^i[n])$, $\forall n$, solve problem (P1.2) and obtain the current optimal solution $(\mathbf{\tilde{q}}_a^*[n],\tilde{P}_a^*[n],\tilde{R}_b^*[n],\tilde{u}_1^*[n])$, $\forall n$; $i=i+1$.}
\STATE{Update the feasible points $(\mathbf{\tilde{q}}_a^i[n]\!,\!\tilde{P}_a^i[n]\!,\!\tilde{R}_b^i[n]\!,\!\tilde{u}_1^i[n])\!=\!(\mathbf{\tilde{q}}_a^*[n],\tilde{P}_a^*[n],\tilde{R}_b^*[n],\tilde{u}_1^*[n])$, $\forall n$.}
\STATE {Compute objective value of (P1.1), denoted by $\bar{R}_b^i$.}
\UNTIL {$|\bar{R}_b^i-\bar{R}_b^{i-1}|\leq\epsilon$ is met, where $\epsilon$ denotes the convergence tolerance.}
\end{algorithmic}
\end{algorithm}

We note that it is very critical to selection the initial feasible solutions/points $(\mathbf{\tilde{q}}_a^0[n],\tilde{P}_a^0[n],\tilde{R}_b^0[n],\tilde{u}_1^0[n])$, $\forall n$, which directly determine the feasibility of the optimization problem (P1.2) and the convergence speed of Algorithm~\ref{alg1}.
Practically, we can generate the initial points randomly and then verify their feasibility. If they are infeasible, we have to generate them randomly again. However, for the optimization problem (P1.2), the aforemetioned random method is not applicable, since it is difficult to randomly generate a trajectory of the UAV to satisfy the mobility constraint \eqref{Mob_C}.

In order to achieve feasible initial points $\mathbf{\tilde{q}}_a^0[n]$, we determine an initial trajectory for the UAV as follows. UAV first flies to Bob from the initial location with the shortest path and  maximum speed, then UAV hovers above Bob's estimated location as long as possible for achieving a better communication channel, and finally UAV flies to the final location with the shortest path and maximum speed in order to arrive at the final location by the end of the last time slot. We note that if the flight period $T$ is not sufficiently large for the UAV to arrive at the location just above Bob and then turn back to the final location, the UAV will turn at a certain midway point and fly to the final location at the end of the last time slot. We refer this initial trajectory as the `line-segment trajectory'. This initial trajectory must satisfy the mobility constraint \eqref{Mob_C}. In addition, we can adjust the UAV's transmit power such that the initial trajectory satisfies the transmission outage probability constraint and the covertness constraint. As such, the generated initial trajectory must be a feasible solution to the optimization problem (P1.2).

For the initial feasible points $\tilde{P}_a^0[n]$ and $\tilde{R}_b^0[n]$, $\forall n$, we can obtain them by randomly generating a very small set of values. Consequently, the initial feasible points of $\tilde{u}_1^0[n]$ can be generated according to \eqref{Det3b}, which is given by
\begin{align}\label{Det13}
\tilde{u}_1^0[n]=\ln\left(1+\frac{\beta_1\tilde{P}_a[n]}{\frac{1}{\varepsilon_w^2}\|\mathbf{\tilde{q}}_a[n]-\mathbf{\hat{q}}_w\|^2+2+\frac{H^2}{\varepsilon_w^2}}\right),~\forall n.
\end{align}

\section{numerical results}

In this section, we provide numerical results to evaluate the performance of the UAV's covert communications achieved by our developed algorithm. To demonstrate the benefit of the joint optimization of the UAV's trajectory and transmit power (denoted as the JTP scheme), we compare it with a benchmark scheme named as the STP scheme, where the UAV's trajectory is fixed and only its transmit power is optimized. Specifically, in the  STP scheme, we adopt the initial feasible trajectory (i.e., the line-segment trajectory detailed in Section IV-D) as its fixed trajectory and then optimize its transmit power accordingly.
Without other statements, the system parameters are set as follows: $\bar{P}_a=20~\mathrm{dBm}$, $P_{a,\max}=4\bar{P}_{a,\max}$, $\varepsilon_b^2=\varepsilon_w^2=25$, $H=100~\mathrm{m}$, $V_{\max}=5~\mathrm{m/s}$,
$\mathbf{q}_{a,0}=[-500,100]^T$,  $\mathbf{q}_{a,F}=[500,100]^T$, $\mathbf{\hat{q}}_w=[100,300]^T$, $\mathbf{\hat{q}}_b=[-100,300]^T$, $\beta_0=-60~\mathrm{dB}$, $\gamma_0=80~\mathrm{dB}$, $\delta_t=1~\mathrm{s}$, $\varrho_{\mathrm{dB}}=3$, $\check{\sigma}_{\mathrm{dB}}^2=-120$, $\rho_b=\rho_w=0.05$, and $\epsilon=10^{-4}$.

\begin{figure}[!t]
  \centering
  \includegraphics[width=3.5in, height=2.8in]{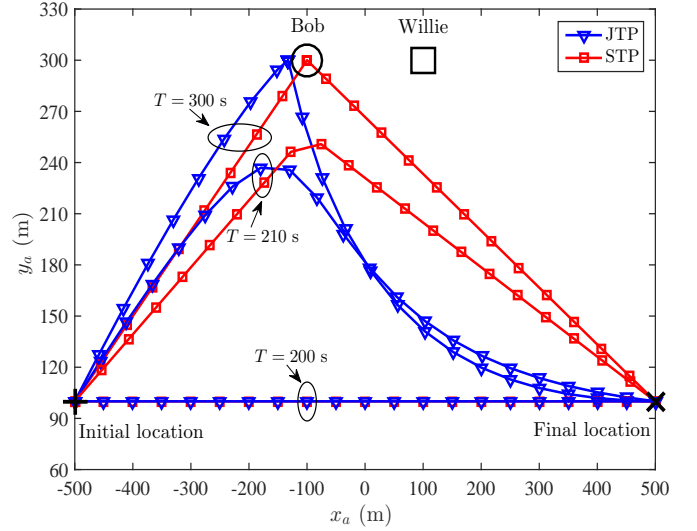}
  \caption{UAV's trajectories for different values of the flight period $T$.}\label{Trajectory}
\end{figure}

\begin{figure}[!t]
  \centering
  \includegraphics[width=3.49in, height=2.8in]{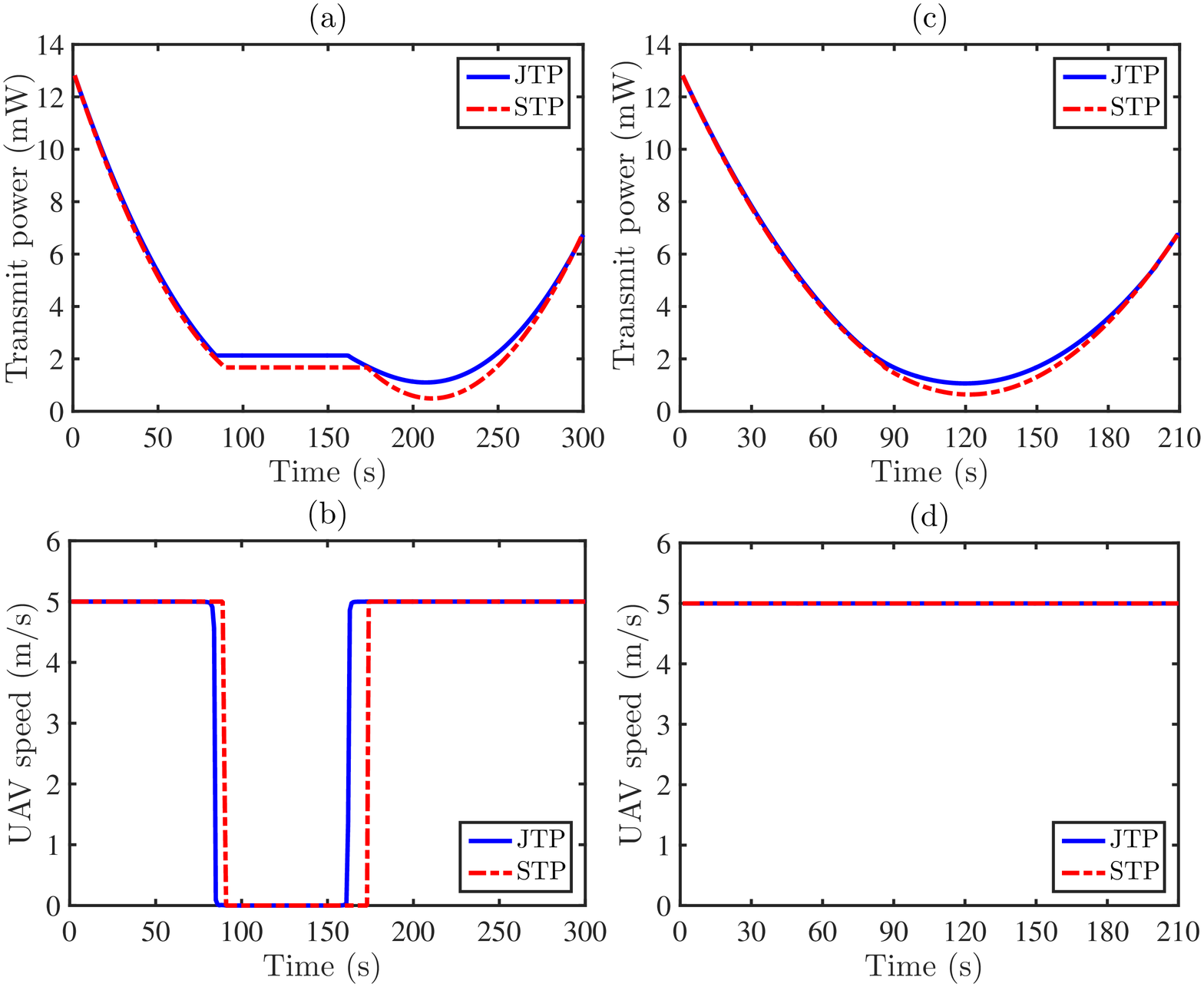}
  \caption{The UAV's transmit power and speed for different values of the flight period $T$, where $T = 300~\mathrm{s}$ for (a) and (b), while $T = 210~\mathrm{s}$ for (c) and (d).}\label{Power_speed}
\end{figure}

In Fig.~\ref{Trajectory}, we plot the optimized trajectories of the UAV achieved by our proposed JTP scheme and the benchmark scheme (i.e., the STP scheme) with different values of the flight period $T$, where Bob's estimated location, Willie's estimated location, the UAV's initial location, and the UAV's final location are marked with $\bigcirc$, $\square$, $+$, and $\times$, respectively.
In this figure, as expected we first observe that the UAV's trajectories achieved by the JTP and STP schemes are identical for $T=200~\mathrm{s}$. This is due to the fact that $T=200~\mathrm{s}$ is the minimum required time for the UAV to fly from its initial location to its final location due to the maximum speed constraint.
However, the UAV's trajectories achieved by these two schemes gradually appear different from each other as $T$ increases.
For example, for $T=300~\mathrm{s}$, in the JTP scheme the UAV first flies to Bob with its maximum speed when it is on the LHS of Bob, then hovers around at Bob for a certain period, and finally chooses a certain path, which is not the straight line from Bob to the UAV's final location but a curve avoiding the closet of Willie, to fly to its final location with the maximum speed at the end of the last time slot. This observation is confirmed by the flight speed of the UAV, which is shown in Fig.~\ref{Power_speed}(b). From this figure, we can see that the above detailed trajectory is significantly different from the line-segment trajectory used in the STP scheme.
Intuitively, the reasons why the UAV hovers around at Bob for a certain period are given as below. When UAV flies from the initial location to the final location, there must be an optimal location that
can achieve the maximum ACTR while satisfying the covertness constraint. As such, UAV prefers to hover this location in order to improve the ACTR when $T$ is sufficiently large. We also
note that the hovering location is near the LHS of Bob, which not only guarantees
a better communication channel but also makes the covertness constraint easier to satisfy.
As such, the hovering location is generally to strike a tradeoff between the communication
performance from the UAV to Bob and average minimum total error rate at Willie.
As we confirmed numerically, this hovering location is the UAV's optimal location in the static scenario.

\begin{figure}[!t]
  \centering
  \includegraphics[width=3.49in, height=2.8in]{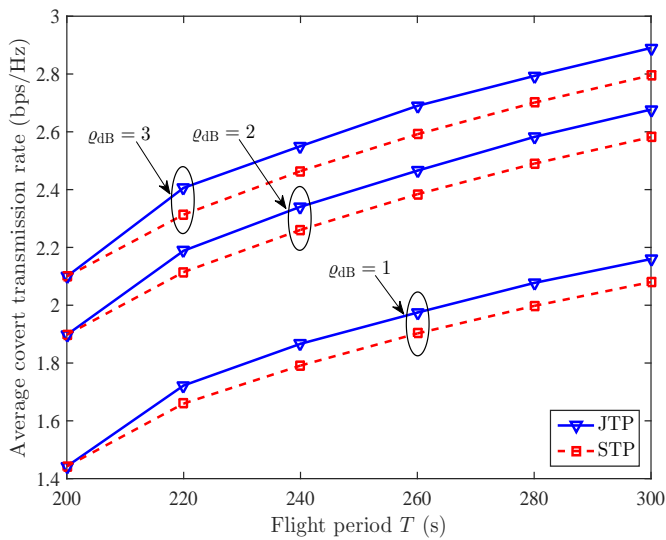}
  \caption{The maximum ACTRs (average covert transmission rates) achieved by the JTP and STP schemes versus the flight period $T$ for different noise uncertainty levels (i.e., different values of $\varrho_{\mathrm{dB}}$).}\label{Rate_T}
\end{figure}

In Fig.~\ref{Power_speed}, we plot the obtained UAV's transmit power and flight speed for different values of the flight period $T$. In Fig.~\ref{Power_speed}(a), we observe that the UAV's transmit power achieved by the JTP scheme is larger than that achieved by the STP scheme.
This is due to the fact that the UAV's trajectory achieved by the JTP scheme is always further away from Willie than that achieved by the STP scheme.
This allows the UAV to transmit signals with a higher transmit power to further improve the ACTR (average covert transmission rate) in the JTP scheme, while ensuring the same level covertness (guaranteeing the same level of detection error rate at Willie). This is confirmed by Fig.~\ref{Rate_T}.
In Fig.~\ref{Power_speed}(a), we also observe that the UAV's transmit power is in three different stages, i.e., rapid-decreasing stage, stable stage, and varying stage. This exactly matches with the UAV's flight trajectory as shown in Fig.~\ref{Trajectory}, i.e., the UAV varies its transmit power as per its location in order to ensure a certain covertness.
In Fig.~\ref{Power_speed}(b), we can see that the UAV's speed at the balancing location is zero, which means that the UAV stays static at this location for a certain time period. In Fig.~\ref{Power_speed}(d), we observe that the UAV always flies at its maximum speed, which is due to the fact that the flight period $T$ is not sufficiently large for the UAV to arrive at Bob's location. This is the reason why the UAV's transmit power as shown in Fig.~\ref{Power_speed}(c) first decreases and then increases.

\begin{figure}[!t]
  \centering
  \includegraphics[width=3.49in, height=2.8in]{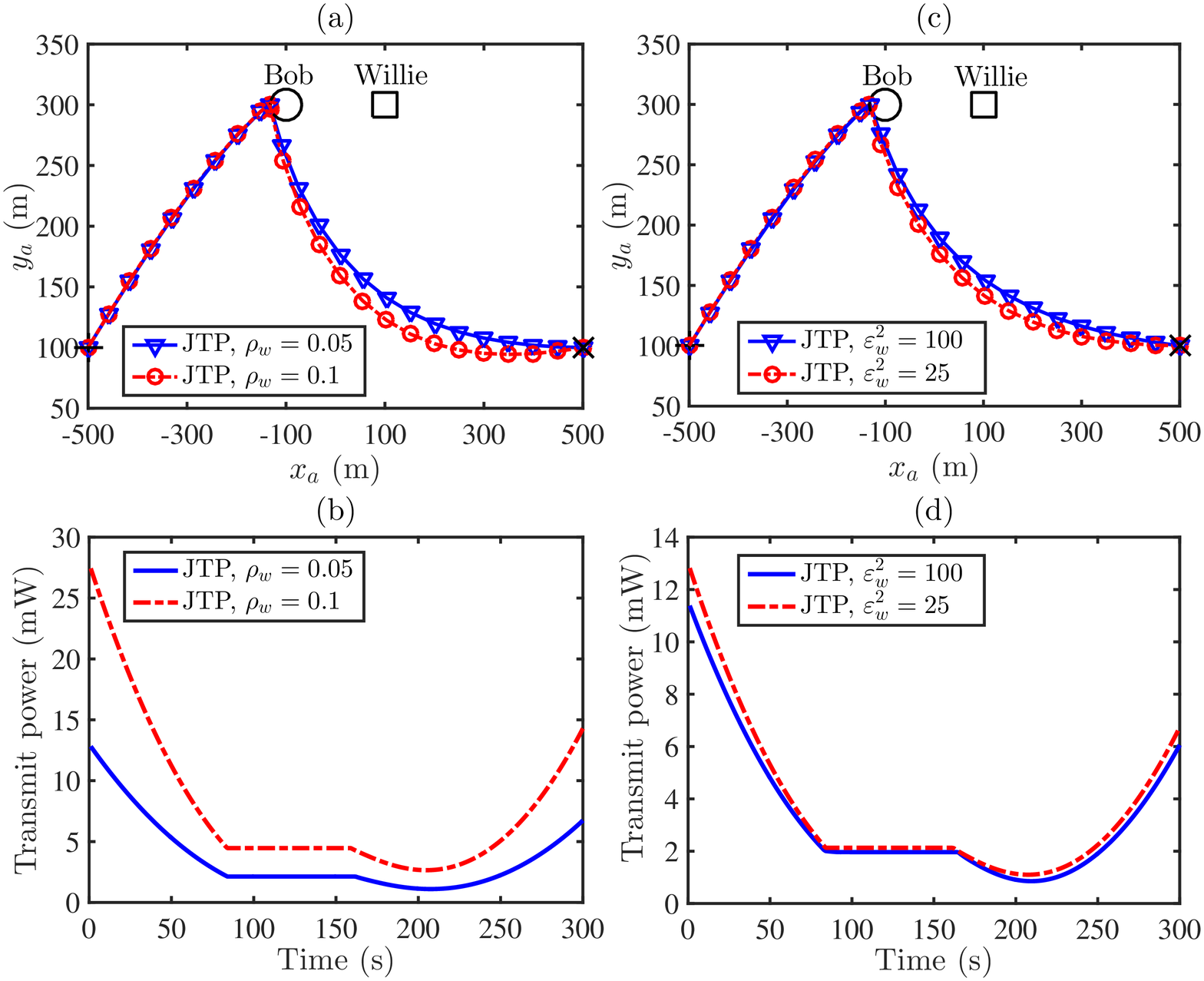}
  \caption{The trajectory of UAV and corresponding transmit power for different  $\rho_w$ and Willie's location uncertainty levels $\varepsilon_w^2$.}\label{Trajectory_Pow_rho_w}
\end{figure}

In Fig.~\ref{Trajectory_Pow_rho_w}, we plot the UAV's trajectory and transmit power achieved by our JTP scheme for different values of the covertness level parameter $\rho_w$ and the location uncertainty parameter $\varepsilon_w^2$.
In Fig.~\ref{Trajectory_Pow_rho_w}(a), we observe that the distance from the UAV's trajectory (from the balancing location to the final location) to Willie decreases as the covertness constraint becomes stricter (i.e., $\rho_w$ decreases in $\check{\xi}^*[n]\geq 1-\rho_w$).
Intuitively, we may expect that the UAV's trajectory should be further away from Willie as the covertness constraint becomes stricter, which is contradict to the above observation. This is due to the fact that the UAV's trajectory and its transmit power are jointly optimized in the JTP scheme. Specifically, the UAV's transmit power has a larger impact on the considered covert communications. As such, for $\rho_w=0.1$, the UAV may choose a trajectory that is further away from the Willie relative to the case with $\rho_w=0.05$, which allows the UAV to transmit with a higher power.
This explanation is confirmed by the observation shown in Fig.~\ref{Trajectory_Pow_rho_w}(b), which shows that the UAV's transmit power for $\rho_w=0.1$ is much larger than that for $\rho_w=0.05$.
In Fig.~\ref{Trajectory_Pow_rho_w}(c) and Fig.~\ref{Trajectory_Pow_rho_w}(d), we observe that the UAV's flight trajectory moves away from Willie and its transmit power increases, as Willie's location uncertainty decreases (i.e., $\varepsilon_w^2$ decreases). We explain this observation by considering a special case. Specifically, as $\varepsilon_w^2\rightarrow\infty$, i.e., the UAV does not know the location information of Willie, the UAV will hover as long as possible around Bob with a small transmit power to ensure the covertness constraint and  chooses the shortest path to fly from the hovering location to its final location. As such, in the specific settings (Willie is on the RHS of the path from Bob to the UAV's final location), we have the aforementioned observation that the UAV's trajectory moves away from Wille as $\varepsilon_w^2$ decreases.

In Fig.~\ref{Rate_T}, we plot the maximum ACTRs achieved by the JTP and STP schemes versus the flight period $T$ for different levels of the noise uncertainty at Willie (i.e., different values of $\varrho_{\mathrm{dB}}$). In this figure, we first observe that the achieved maximum ACTRs by the two schemes increase with $T$. This is due to the fact that the covert transmission rates are relatively large when the UAV is hovering around Bob and a larger $T$ allows the UAV to hover around Bob for a longer time. In this figure, we also observe that the proposed JTP scheme always achieves a higher maximum ACTR than the STP scheme, which demonstrates that the joint optimization of the UAV's trajectory and transmit power is necessary for improving the covert communication performance in the considered UAV networks. Finally, as expected we observe that this maximum ACTR increases as the noise uncertainty at Willie increases (i.e., as $\varrho_{\mathrm{dB}}$ increases). This is due to the fact that a larger $\varrho_{\mathrm{dB}}$ makes it harder for Willie to make correct decisions in the detection of the UAV's covert transmission.

\begin{figure}[!t]
  \centering
  \includegraphics[width=3.49in, height=2.8in]{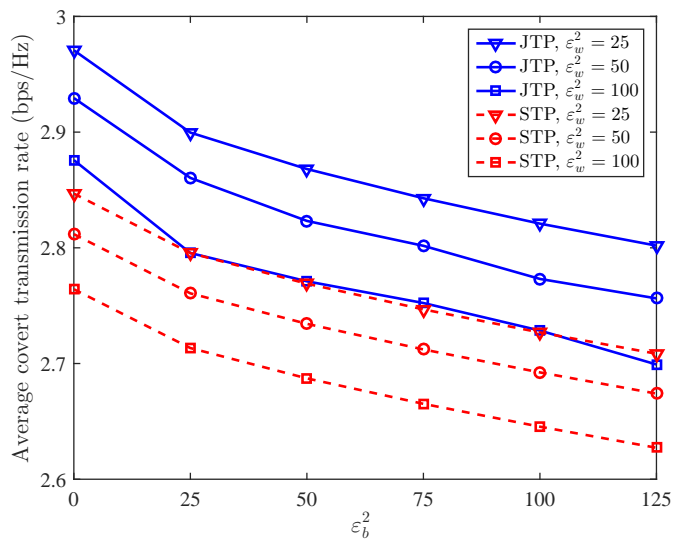}
  \caption{The maximum ACTRs (average covert transmission rates) achieved by the JTP and STP schemes versus the Bob's location uncertainty parameter $\varepsilon_b^2$ for different values of Willie's location uncertainty parameter $\varepsilon_w^2$.}\label{Rate_ksi_b_ksi_w}
\end{figure}

In Fig.~\ref{Rate_ksi_b_ksi_w}, we plot the maximum ACTRs achieved by the JTP and STP schemes versus Bob's location uncertainty level $\varepsilon_b^2$ for different values of Willie's location uncertainty $\varepsilon_w^2$. In this figure, as expected we observe that the maximum ACTR decreases as $\varepsilon_b^2$ or $\varepsilon_w^2$ increases, since a larger $\varepsilon_b^2$ makes the transmission outage constraint from the UAV to Bob harder to be satisfied, while a larger $\varepsilon_w^2$ makes the covertness constraint more difficult to satisfy. This observation also shows that Bob or Willie's location uncertainty has a large impact on the achieved maximum ACTR.


\section{Conclusion}

In this work, we jointly optimized the UAV's trajectory and transmit power for covert communications, in terms of maximizing the ACTR (average covert transmission rate) from the UAV to Bob subject to the transmission outage and covertness constraints. The formulated optimization problem serves as a new design framework to jointly consider the transmission rate from the UAV to Bob and the detection error rate at Willie, which have been transformed into convex forms by utilizing first-order restrictive approximation in this work. This allows us to develop an algorithm based on SCA to solve the formulated optimization problem.
Our examinations showed that the developed joint optimization of the UAV's trajectory and transmit power achieves significantly better covert communication performance relative to a benchmark scheme. Interesting, we found that, with regard to the UAV's trajectory, it prefers to hover around Bob for a certain period and, surprisingly, it moves closer to Willie as the covertness becomes stricter.

\appendices
\section{Proof of Lemma 1}
 As per \eqref{False} and \eqref{Miss}, the total error rate $\xi[n]$ at the $n$-th time slot is given by
\begin{align}\label{Appa_1}
\xi[n]&=
\begin{cases}
1,&P_{th}[n]< \frac{\check{\sigma}^2}{\varrho},\\
\tau_1[n],& \frac{\check{\sigma}^2}{\varrho}\leq P_{th}[n]< E_w[n]+\frac{\check{\sigma}^2}{\varrho},\\
\tau_1[n]+\tau_2[n],&E_w[n]+\frac{\check{\sigma}^2}{\varrho}\leq P_{th}[n]\leq{\varrho\check{\sigma}^2},\\
\tau_2[n],& {\varrho\check{\sigma}^2}<P_{th}[n]\leq E_w[n]+{\varrho\check{\sigma}^2},\\
1,& P_{th}[n]>E_w[n]+{\varrho\check{\sigma}^2},
\end{cases}
\end{align}
where
\begin{align}\label{Appa_2}
&\tau_1[n]=\frac{1}{2\ln{(\varrho)}}\ln{\left(\frac{\varrho\check{\sigma}^2}{P_{th}[n]}\right)},\tau_2[n]=\frac{\ln{\left(\frac{\varrho(P_{th}[n]-E_w[n])}{\check{\sigma}^2}\right)}}{2\ln{(\varrho)}},\nonumber\\
&\tau_1[n]+\tau_2[n]=\frac{1}{2\ln{(\varrho)}}\ln{\left(\varrho^2-\frac{\varrho^2E_w[n]}{P_{th}[n]}\right)}.
\end{align}
We first note that $\xi[n] = 1$ is the worst-case scenario for Willie and thus Willie does not set $P_{th}[n]<\frac{\check{\sigma}^2}{\varrho}$ or $P_{th}[n]>E_w[n]+{\varrho\check{\sigma}^2}$. We also oberve that $\xi[n]$ monotonically decreases with $P_{th}[n]$ for $\frac{\check{\sigma}^2}{\varrho}\leq P_{th}[n]< E_w[n]+\frac{\check{\sigma}^2}{\varrho}$, while it is a monotonically increasing function of $P_{th}[n]$ when $E_w[n]+\frac{\check{\sigma}^2}{\varrho}\leq P_{th}[n]\leq{\varrho\check{\sigma}^2}$ and ${\varrho\check{\sigma}^2}<P_{th}[n]\leq E_w[n]+{\varrho\check{\sigma}^2}$.
Furthermore, it is easy to verify that $\xi[n]$ is a continuous function of $P_{th}[n]$ in \eqref{Appa_1}. As such, we can conclude that the optimal detection threshold is given by $P_{th}^*[n]=E_w[n]+\frac{\check{\sigma}^2}{\varrho}$ and the corresponding minimum total error rate is given by $\xi^*[n]=\frac{1}{2\ln{(\varrho)}}\ln{\left(\frac{\varrho\check{\sigma}^2}{E_w[n]+\frac{\check{\sigma}^2}{\varrho}}\right)}$. This completes the proof of Lemma~1.\hfill$\blacksquare$
\newcounter{mytempeqncnt1}
\begin{figure*}[!ht]
\normalsize
\setcounter{mytempeqncnt1}{\value{equation}}
\setcounter{equation}{59}
\begin{subequations}\label{Appd_1}
\begin{align}
&\mathbb{E}_{X[n]}\left[g(X[n])\right]\leq g(\lambda[n]+2)+\sqrt{\frac{2}{\pi}}\sqrt{\lambda[n]+1}\left(\lim_{X[n]\rightarrow\infty}\frac{g(X[n])}{X[n]}+\frac{g(0)-g(\lambda[n]+2)}{\lambda[n]+2}\right)\label{Appd_1a}\\
&=\ln\left(1+\frac{\beta_1P_a[n]}{\lambda[n]+2+\frac{H^2}{\varepsilon_w^2}}\right)+\frac{\sqrt{\frac{2}{\pi}}\sqrt{\lambda[n]+1}}{\lambda[n]+2}\left(\ln\left(1+\frac{\beta_1\varepsilon_w^2P_a[n]}{H^2}\right)-\ln\left(1+\frac{\beta_1P_a[n]}{\lambda[n]+2+\frac{H^2}{\varepsilon_w^2}}\right)\right)\label{Appd_1b}\\
&\leq \ln\left(1\!+\!\frac{\beta_1P_a[n]}{\lambda[n]\!+\!2\!+\!\frac{H^2}{\varepsilon_w^2}}\right)\!+\!\frac{\sqrt{2}}{\sqrt{\pi}\sqrt{\lambda[n]+1}}\left(\ln\left(1\!+\!\frac{\beta_1\varepsilon_w^2 P_a[n]}{H^2}\right)\!-\!\ln\left(1+\frac{\beta_1P_a[n]}{\lambda[n]\!+\!2\!+\!\frac{H^2}{\varepsilon_w^2}}\right)\right)\triangleq g^u(\lambda[n],P_a[n]).\label{Appd_1c}
\end{align}
\end{subequations}
\setcounter{equation}{\value{mytempeqncnt1}}
\hrulefill
\vspace*{4pt}
\end{figure*}
\section{Proof of Lemma 2}
Following \eqref{XN}, the random variable $X[n]$ can be written as
\begin{align}\label{Appb_1}
X[n]\!=\!\frac{1}{\varepsilon_w^2}[(x_a[n]\!-\!\hat{x}_w\!-\!\Delta x_w)^2\!+\!(y_a[n]\!-\!\hat{y}_w\!-\!\Delta y_w)^2].
\end{align}
As mentioned before, the Gaussian error model is adopted in the considered scenario, where $\Delta x_w$ and $\Delta y_w$ follow the same distribution $\mathbb{N}(0,\varepsilon_w^2)$. Then, we have
\begin{align}\label{Appb_2}
\frac{1}{\varepsilon_w}(x_a[n]-\hat{x}_w-\Delta x_w)\sim\mathbb{N}(\frac{x_a[n]-\hat{x}_w}{\varepsilon_w},1),~\forall n,\nonumber\\
\frac{1}{\varepsilon_w}(y_a[n]-\hat{y}_w-\Delta y_w)\sim\mathbb{N}(\frac{y_a[n]-\hat{y}_w}{\varepsilon_w},1),~\forall n.
\end{align}
For the random variable $x=\sum_{i=1}^vx_i^2$, where $x_i$, $i=1,\cdots, v$, are  i.i.d. random variables with $x_i\sim \mathbb{N}(\mu_i,1)$, $\forall i$, $x$ follows the noncentral chi-square distribution with $v$ degrees of freedom and with the noncentral parameter $\sum_{i=1}^v\mu_i^2$.
As such, the random variable $X[n]$ follows the noncentral chi-square distribution with $2$ degrees of freedom and with the noncentral parameter $\lambda[n]$, which is given by
\begin{align}\label{Appb_3}
&\lambda[n]\triangleq \frac{1}{\varepsilon_w^2}\Big((x_a[n]-\hat{x}_w)^2+(y_a[n]-\hat{y}_w)^2\Big),~\forall n,
\end{align}
and the expectation and variance of the $X[n]$ are given by $\lambda[n]+2$ and $4\lambda[n]+4$, respectively.
This completes the proof of Lemma~2.\hfill$\blacksquare$

\section{Proof of Lemma 4}
We note that $v_i\sim\mathbb{N}(\mu_i,1)$, $i=1,2,\cdots,k$.
 For the random variable $V=\sum_{i=1}^kv_i^2$, its noncentral parameter is given by $\eta=\sum_{i=1}^k\mu_i^2$. When $k\rightarrow\infty$, we can achieve the result in Lemma $4$ by using the Central Limit Theorem. In the following, we prove that Lemma~4 still holds for the case where the degree of freedom $k$ is small but the noncentral parameter $\eta$ is sufficiently large.

It can be observed from \eqref{PDF_x} that the pdf of a noncentral chi-square random variable is only related to the noncentral parameter $\eta$ and is independent of individual mean $\mu_i$, $i=1,2,\cdots,k$. Following this observation, we introduce a random variable $S$, which is defined as
\begin{align}\label{Appc_1}
S\triangleq \left(s_1+\sqrt{\eta}\right)^2+\sum_{i=2}^ks_i^2,
\end{align}
where $s_i$, $\forall i$, follows $\mathbb{N}(0,1)$. Then, the random variable $S$
must follow the noncentral chi-square distribution with $k$ degrees of freedom and with $\eta$ as the noncentral parameter. As such, $V$ and $S$ are i.i.d. random variables. In the following, we aim to prove that the random variable $S$ can be well approximated as a Gaussian random variable when the degree
of freedom $k$ is small but the noncentral parameter $\eta$ is sufficiently large. To this end, we rearrange $S$ as
\begin{align}\label{Appc_2}
S=\sum_{i=1}^ks_i^2+2s_1\sqrt{\eta}+\eta.
\end{align}
We note that the term $\sum_{i=1}^ks_i^2$ in \eqref{Appc_2} follows the (central) chi-square distribution with mean $k$ and variance $2k$, and
\begin{align}\label{Appc_3}
2s_1\sqrt{\eta}+\eta \sim \mathbb{N}(\eta,4\eta).
\end{align}
We also note that, if $k$ is small and $\eta$ is large enough, the distribution of the random variable $S$ is dominated by the Gaussian distribution. As such, the pdf of the noncentral chi-square random variable $S$ can be well approximated by the Gaussian distribution $\mathbb{N}(k+\eta,2k+4\eta)$.
This completes the proof of Lemma~4.\hfill$\blacksquare$

\section{Proof of Proposition 1}
Following Lemma~4, the random variable $X[n]$ can be well approximated as a Gaussian random variable with mean $2+\lambda[n]$ and variance $4+4\lambda[n]$.
According to Lemma~3, the mean absolute deviation of the random variable $X[n]$ is given by $2\sqrt{\frac{2}{\pi}}\sqrt{\lambda[n]+1}$. In addition, we note that $g(X[n])$ is a convex function with respect to $X[n]$ and $\lim_{X[n]\rightarrow\infty}\frac{g(X[n])}{X[n]}=0$. As such, following \eqref{lemma2_2}, an upper bound on $\mathbb{E}_{X[n]}\left[g(X[n])\right]$ is given by \eqref{Appd_1b}, which is shown at the top of the previous page. Considering that $\lambda[n]$ is large, \eqref{Appd_1b} can be further simplified into \eqref{Appd_1c}.
As such, a lower bound on the average minimum
total error rate $\bar{\xi}^*[n]$ is given by
\setcounter{equation}{60}
\begin{align}\label{Appd_2}
\bar{\xi}^*[n]\geq 1-\frac{1}{2\ln{(\varrho)}}g^u(\lambda[n],P_a[n]),
\end{align}
where $g^u(\lambda[n],P_a[n])$ is given in \eqref{Appd_1c}.
This completes the proof of Proposition 1.\hfill$\blacksquare$

\section{Proof of Lemma 6}
We first define $f(x_1,x_2)\triangleq\frac{1}{x_1x_2}$, where $x_1>0$ and $x_2>0$. The Hessian matrix of $f(x_1,x_2)$ is given by
\begin{align}\label{Appe_1}
&\triangledown^2f(x_1,x_2)=\frac{1}{x_1x_2}
\left[\begin{matrix}
   \frac{2}{x_1^2} & \frac{1}{x_1x_2}  \\
   \frac{1}{x_1x_2} & \frac{2}{x_2^2}
\end{matrix}\right]\nonumber\\
&=\frac{1}{x_1x_2}\left[
\begin{matrix}
   \frac{1}{x_1^2} & 0  \\
   0 & \frac{1}{x_2^2}
\end{matrix}\right]
+\frac{1}{x_1x_2}
\left[\begin{matrix}
   \frac{1}{x_1} \\
   \frac{1}{x_2}
\end{matrix}\right]
\left[\begin{matrix}
   \frac{1}{x_1} &\frac{1}{x_2}
\end{matrix}\right]\succeq \mathbf{0}.
\end{align}
As such, $f(x_1,x_2)$ is a convex function with respect to $x_1$ and $x_2$ when $x_1>0$ and $x_2>0$. We note that a convex function is lower bounded by its first-order approximation. Consequently, for given feasible points $\tilde{x}_1$ and $\tilde{x}_2$, the following inequality must hold
\begin{align}\label{Appe_2}
\frac{1}{x_1x_2}\geq \frac{1}{\tilde{x}_1\tilde{x}_2}-\frac{(x_1-\tilde{x}_1)}{\tilde{x}_2\tilde{x}_1^2}-\frac{(x_2-\tilde{x}_2)}{\tilde{x}_1\tilde{x}_2^2}.
\end{align}

We note that the term $\frac{P_a[n]\gamma_0}{2^{R_b[n]}-1}$ can be rearranged as $\frac{\gamma_0}{(P_a[n])^{-1}(2^{R_b[n]}-1)}$. According to \eqref{Appe_1}, it is easy to verify that $\frac{\gamma_0}{(P_a[n])^{-1}(2^{R_b[n]}-1)}$ is jointly convex with respect to $(P_a[n])^{-1}$ and $2^{R_b[n]}-1$. As such, replacing $x_1$, $x_2$, $\tilde{x}_1$, and $\tilde{x}_2$ in \eqref{Appe_2} with $(P_a[n])^{-1}$, $2^{R_b[n]}-1$, $(\tilde{P}_a[n])^{-1}$, and $2^{\tilde{R}_b[n]}-1$, respectively, we can obtain the inequality \eqref{Pout5}. This completes the proof of Lemma~6.\hfill$\blacksquare$

\section{Proof of Lemma 7}
We first define $f(x_1,x_2)\triangleq\sqrt{x_1x_2}$, where $x_1>0$ and $x_2>0$. The Hessian matrix of $f(x_1,x_2)$ is given by
\begin{align}\label{Appf_1}
&\triangledown^2f(x_1,x_2)=-\frac{1}{4}\sqrt{x_1x_2}
\left[\begin{matrix}
   \frac{1}{x_1} \\
   \frac{-1}{x_2}
\end{matrix}\right]
\left[\begin{matrix}
   \frac{1}{x_1} &\frac{-1}{x_2}
\end{matrix}\right]\preceq\mathbf{0}.
\end{align}
As such, $f(x_1,x_2)$ is jointly concave with respect to $x_1$ and $x_2$ when $x_1>0$ and $x_2>0$. We recall that the first-order approximation of any concave function is its upper bound. Consequently, for given feasible points $\tilde{x}_1$ and $\tilde{x}_2$, the following inequality must hold
\begin{align}\label{Appf_2}
\sqrt{x_1x_2}\leq \sqrt{\tilde{x}_1\tilde{x}_2}\!+\!\frac{(x_1-\tilde{x}_1)}{2}\sqrt{\frac{\tilde{x}_2}{\tilde{x}_1}}\!+\!\frac{(x_2-\tilde{x}_2)}{2}\sqrt{\frac{\tilde{x}_1}{\tilde{x}_2}}.
\end{align}

We note that $\frac{1}{\varepsilon_w^2}\|\mathbf{q}_a[n]-\mathbf{\hat{q}}_w\|^2+1>0$ and $u_1[n]>0$ must hold. Following \eqref{Appf_1}, $\sqrt{\big(\frac{1}{\varepsilon_w^2}\|\mathbf{q}_a[n]-\mathbf{\hat{q}}_w\|^2+1\big)u_1[n]}$ is jointly concave with respect to $\frac{1}{\varepsilon_w^2}\|\mathbf{q}_a[n]-\mathbf{\hat{q}}_w\|^2+1$ and $u_1[n]$. Replacing $x_1$, $x_2$, $\tilde{x}_1$, and $\tilde{x}_2$ in \eqref{Appf_2} with $\frac{1}{\varepsilon_w^2}\|\mathbf{q}_a[n]-\mathbf{\hat{q}}_w\|^2+1$, $u_1[n]$, $\frac{1}{\varepsilon_w^2}\|\mathbf{\tilde{q}}_a[n]-\mathbf{\hat{q}}_w\|^2+1$, and $\tilde{u}_1[n]$, respectively, we can obtain the inequality \eqref{Det5}. This completes the proof of Lemma~7.\hfill$\blacksquare$

\bibliographystyle{IEEEtran}
\bibliography{IEEEfull,UAV_Covert}


\ifCLASSOPTIONcaptionsoff
  \newpage
\fi

\end{document}